\documentclass[twocolumn,amssymb, nobibnotes, aps, prxquantum, superscriptaddress]{revtex4-2}

\usepackage{amsmath}
\usepackage{braket}
\usepackage{bm}
\usepackage{graphicx}
\usepackage{float} 
\usepackage{subfigure} 
\usepackage{color}
\usepackage[normalem]{ulem}
\usepackage[ruled,linesnumbered]{algorithm2e}
\usepackage{array}
\usepackage{comment}
\usepackage[utf8]{inputenc}
\usepackage[T1]{fontenc}
\usepackage[english]{babel}
\makeatletter
\addto\captionsenglish{\renewcommand{\selectlanguage}[1]{}}
\makeatother

\usepackage{hyperref}

\begin{document}

\title{AI-enhanced Quantum Simulation of Schwinger Model}

\author{Ao-Ning Wang}
\affiliation{Department of Physics and HK Institute of Quantum Science \& Technology,
The University of Hong Kong, Pokfulam Road, Hong Kong, China}
\affiliation{Hong Kong Branch for Quantum Science Center of Guangdong-Hong Kong-Macau Greater Bay Area, Shenzhen, China}
\author{Min-Quan He}
\email{hemq@hku.hk}
\affiliation{Department of Physics and HK Institute of Quantum Science \& Technology,
The University of Hong Kong, Pokfulam Road, Hong Kong, China}
\affiliation{Hong Kong Branch for Quantum Science Center of Guangdong-Hong Kong-Macau Greater Bay Area, Shenzhen, China}
\author{Z. D. Wang}
\email{zwang@hku.hk}
\affiliation{Department of Physics and HK Institute of Quantum Science \& Technology,
The University of Hong Kong, Pokfulam Road, Hong Kong, China}
\affiliation{Hong Kong Branch for Quantum Science Center of Guangdong-Hong Kong-Macau Greater Bay Area, Shenzhen, China}

\begin{abstract}
    The Schwinger Model from Quantum Electrodynamics~(QED) has long served as a valuable simplified model for exploring key physical phenomena in Quantum Chromodynamics~(QCD)—a field rich with fundamental insights but is substantially more complex. While the phase diagram of the Schwinger Model bears extraordinary significance and remains challenging to investigate, recent progress on the model mainly focuses on detailed case studies. Here, we propose a model that we refer as the Neural Network Facilitated Implicit Quantum Simulation~(NN-IQS) model as a solution. After training on limited discrete data points on the Schwinger Model phase diagram, the NN-IQS model allows quick generation of extra sample points over a continuous domain. The model can even generalize beyond its training range, maintaining robust performance in previously unexplored parameter space and system sizes.
\end{abstract}

\maketitle

\section{Introduction}

\begin{figure*}
    \centering
    \includegraphics[width=1.0\linewidth]{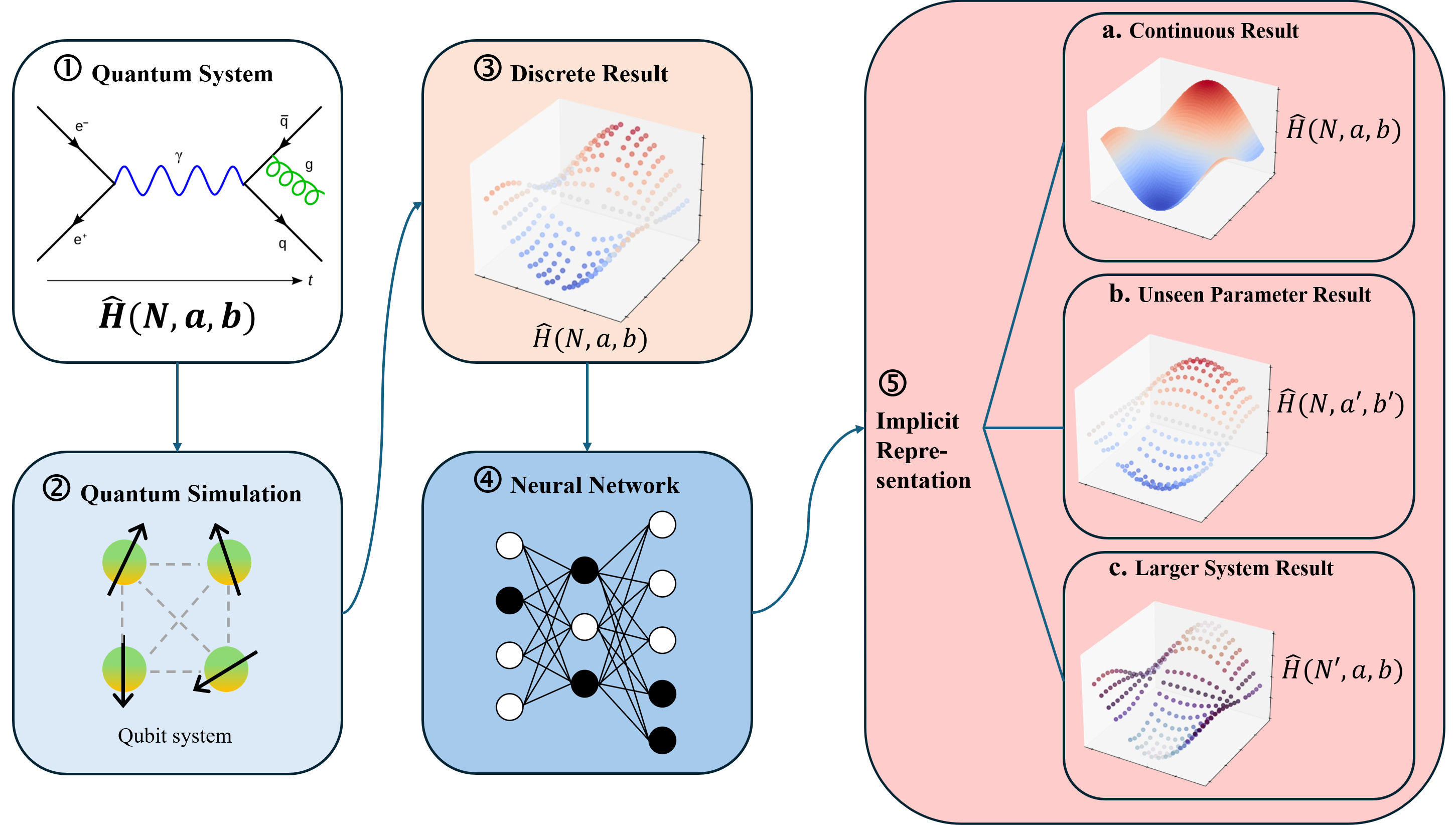}
    \caption{The NN-IQS Model. Its workflow consists of the following steps: 1) select a quantum system of interest and define its Hamiltonian for simulation; 2) perform simulations of the chosen system using either a quantum device or a classical quantum simulator; 3) obtain the discrete simulation result; 4) format and feed the discrete result into the neural network; 5) three possible results of training: a. the discrete result is processed into continuous one; b. predicted results for parameter ranges not included in the training set; c. predicted results for larger system sizes than those used during training. If the required discrete data can be obtained through other sources, such as experimental measurements or alternative computational techniques, the NN-IQS framework can be applied flexibly from step 3).}
    \label{figure_1}
\end{figure*}

The Schwinger Model has been a popular topic in modern high energy physics research as a simplified version of QCD~\cite{schwinger_1, schwinger_2, schwinger_3, schwinger_4, schwinger_5, schwinger_6, schwinger_7, schwinger_8}. While QCD, especially its phase diagram, is widely believed to hold exceptional physical importance, both theoretical~\cite{symmetry1302079, Stephanov2012, LATTICE2021, PHysRevD101054032} and experimental~\cite{heavyion, PHysRevD101054032, Stephanov2012, 200211957v1} studies of it are facing major challenges. The Schwinger Model, on the other hand, bears favorable simplicity and flexibility~\cite{hirokiohata, xiaowei, rossdempsey, yuyashimizu, nbutt, mbanuls, erico, hirotsugufujii, lfuncke, tbyrnes}. Although resides within a different framework (QED), it exhibits high physical similarities to the QCD model~\cite{xiaowei, yuyashimizu, lfuncke}, making it a perfect alternative option for QCD research. Out of various research methods, quantum simulation is a promising numerical tool for studying the Schwinger Model~\cite{xiaowei}, as it is for many other quantum systems~\cite{180805232, 11031399, guerreschi}. By mapping the Schwinger Model into a spin model, scientists can closely monitor its quantum behavior~\cite{11031399}. However, the massive time and computational resources needed for quantum simulation remain an issue due to exponential complexity~\cite{11031399, 1s20s0167, 1s20s001046, physrevx1004}.

Various classical AI algorithms have been raised in recent years to alleviate this issue~\cite{physrevx8031084, scipostphyslectnotes29, photonics0800033, 220700298v3, physreva107010101}, with a typical example of using neural networks to explicitly represent the target quantum system during a quantum simulation procedure~\cite{replearn1, replearn2, replearn3}. Nevertheless, the "implicit" version of representation learning, where the whole quantum system is learned altogether as a functional mapping, like the one case reported in~\cite{221113765v1}, is not well studied. Compared to training a neural network to represent a single, fixed quantum setting, encoding the entire quantum system within the network offers several advantages. Most notably, the model would not require retraining each time the system parameters change, and it would possess greater potential for extension usage.

Here, we propose for the first time, a model that employs a neural network--referred to as the Neural Network Facilitated Implicit Quantum Simulation (NN-IQS) model, to implicitly represent the Schwinger Model, as illustrated in Figure \ref{figure_1}. We demonstrate with numerical results that, with one single round of training on limited discrete data points on the phase diagram of the Schwinger Model, our NN-IQS model can accurately predict the desired physical property at any arbitrary coordinate on the diagram, outperforming all three interpolation benchmarks. The model further exhibits strong extensibility, successfully predicting data in parameter regimes and system sizes outside the training domain. Moreover, it is immediately deployable on current quantum devices and easily adapts to a broad class of quantum systems apart from the Schwinger Model. These features imply our work as a fresh paradigm for addressing quantum problems that are either analytically intractable or computationally expensive. A large number of unsolvable quantum systems with theoretical ambiguity, parameter barriers or scaling hindering may become tangible with this new pathway. Our work also establishes a new quantum–classical resource allocation strategy in the NISQ era, enabling access to complex and hard problems that were previously costly to simulate.

\section{Simulation of Schwinger Model}
The Schwinger Model we employ in this work to reflect on the complex QCD problem is a 1+1D single-flavor massless QED model. The spin Hamiltonian takes the form~\cite{betavqe}:
\begin{equation}
\label{Hamiltonian}
\begin{aligned}
    \hat{H} = \frac{g}{8 w} \sum_{n=2}^{N - 1} \sum_{1 \leq k < l\leq n}Z_{k}Z_{l} + \frac{1}{2} \sum_{n=1}^{N-1} \frac{w}{g} [ X_{n}X_{n+1} + Y_{n}Y_{n+1} ]\\ + \frac{1}{2}\sum_{n=1}^{N}(\frac{m}{g} (-1)^{n} + \frac{\mu}{g}) Z_{n} - \frac{g}{8w} \sum_{n=1}^{N-1} (n \mod 2) \sum_{l=1}^{n} Z_{l},
\end{aligned}
\end{equation}
where $m$ is the mass, $g$ is the dimension-full coupling constant, $\mu$ is the chemical potential, and $w = 1/2a$ is determined by the lattice approximation strategy.

The phase diagram of the Schwinger Model is usually defined as a plot of chiral condensation values across different $T$ and $\mu$ values, where chiral condensation is widely used as an indication of phase transition and will exhibit a zero-to-finite transition at the boundary of symmetry breaking~(if there is any).

To plot the phase diagram, we will directly solve the expected values of chiral condensation. For a fixed set of $T$ and $\mu$, the Gibbs state of the system at finite temperature $T$ is given by $\rho(T) = e^{-H/T} /||e^{-H/T} ||$. The expected chiral condensation is then $\frac{1}{2Na} tr(\rho(T) \sum_{n=1}^{N} (-1)^{n} Z_{n})$~\cite{betavqe,adiabatic}. A detailed derivation of the chiral condensation operator is provided in~\cite{betavqe,adiabatic}, and we attach it in Appendix C.

Theoretically, explicit symmetry breaking in the Schwinger model requires both a finite fermion mass $m$ and a finite topological parameter $\theta$~\cite{schwingertransition}, which increase the model’s complexity and diminish its role as a simplified analogue to QCD.

That is why in our case we adopt the 1+1D single-flavor massless Schwinger model, where fermion mass is set to zero, and the topological parameter is ignored. As a consequence, there will not be explicit phase transitions in the Schwinger Model we use, given that chiral condensation takes a finite value in the whole range of $T$ and $\mu$~\cite{schwingertransition}. We employ a similar approach as A. Tomiya~\cite{betavqe} to define our phase transition in this situation. We will normalize our result and locate the region where chiral condensation is changing dramatically.

After the simulation, we will also cross-check with the theoretical value given in~\cite{theory}, which states that
\begin{equation}
\label{theo_result}
  \left\langle\Bar{\Psi}\Psi\right\rangle = -\frac{m_\gamma}{2\pi}e^{\gamma}e^{2I(\beta m_\gamma)}
\end{equation}
Where $m_\gamma = g/\sqrt{\pi}$, $\beta = 1/T$, $\gamma = 0.57721...$ is the Euler-Mascheroni constant, and the function $I(a)$ satisfies
\begin{equation}
    I(a) = \int_{0}^{\infty}\frac{1}{1 \, - \, e^{acosh(t)}}dt
\end{equation}

\begin{figure}
    \centering
    \includegraphics[width=0.9\linewidth]{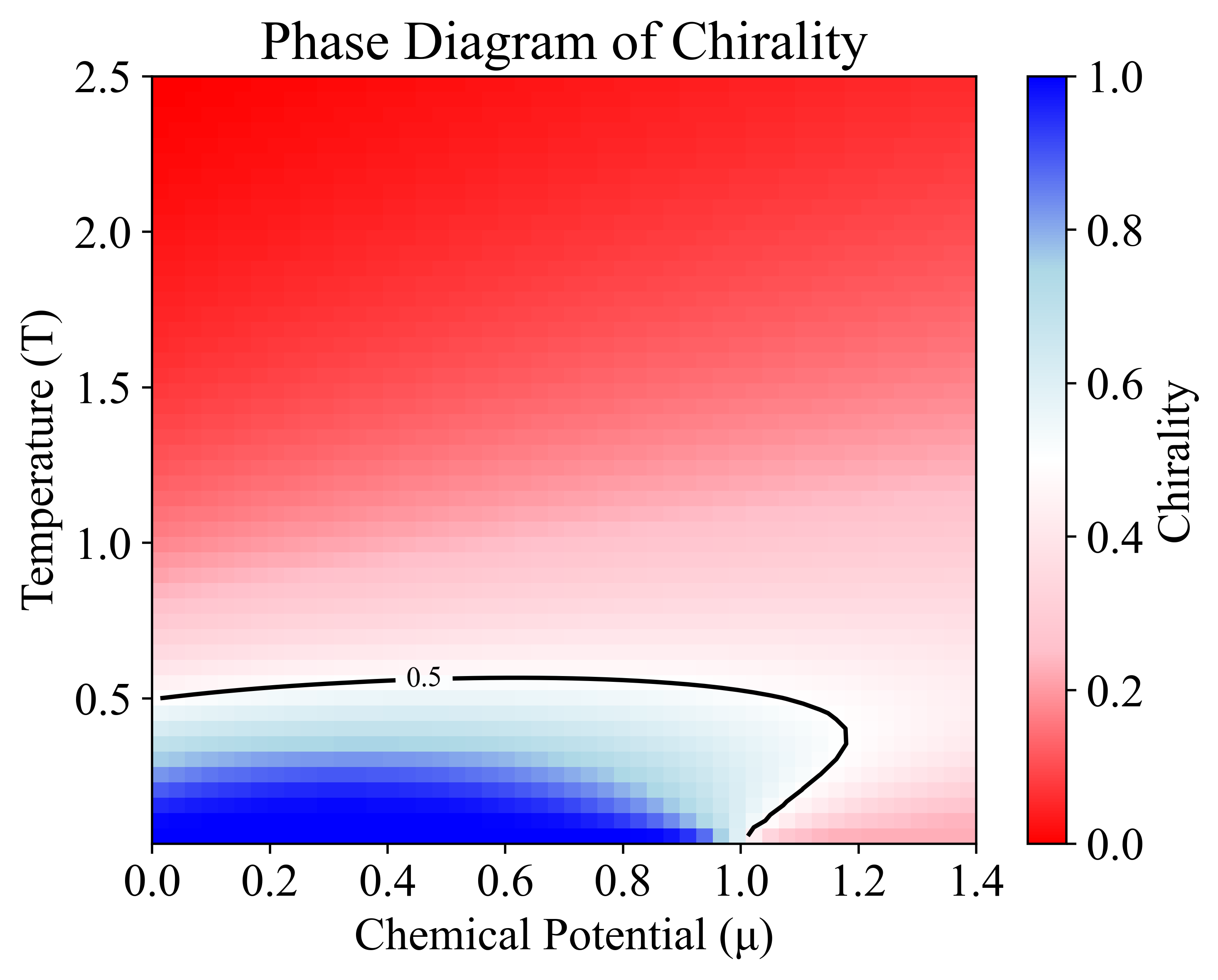}
    \caption{Chiral condensation simulation result, min-max normalized. The 0.5 contour is plotted to mark the approximate location of transition region. $X$ axis is the chemical potential $\mu$ in the system Hamiltonian, while $Y$ axis is the temperature $T$ in Gibbs state. }
    \label{est_chiral}
\end{figure}

To plot the phase diagram of $T\&\mu$, $\mu/g$ is taken to be consecutive values within $[0, 1.4]$, and $T/g$ taken within $[0, 2.5]$, same as that have been done in~\cite{betavqe}, because this is the region where transition occurs. $g/w$ is set to a relatively small value $0.3$ in order to be consistent with the continuous approximation ($ag\rightarrow0$). For illustration, lattice point number~(qubit number) is set to 10. Analytical calculation yields a phase diagram as shown in Figure \ref{est_chiral}, in which all chiral condensation values have been rescaled between $[0,1]$ by min-max normalization. It can be seen clearly that chirality changes dramatically around the normalized value of 0.5, as the 0.5 contour indicates. Therefore, we define the normalized region between 0.4 and 0.6 as our transition region.

\begin{figure}
    \centering
    \includegraphics[width=0.9\linewidth]{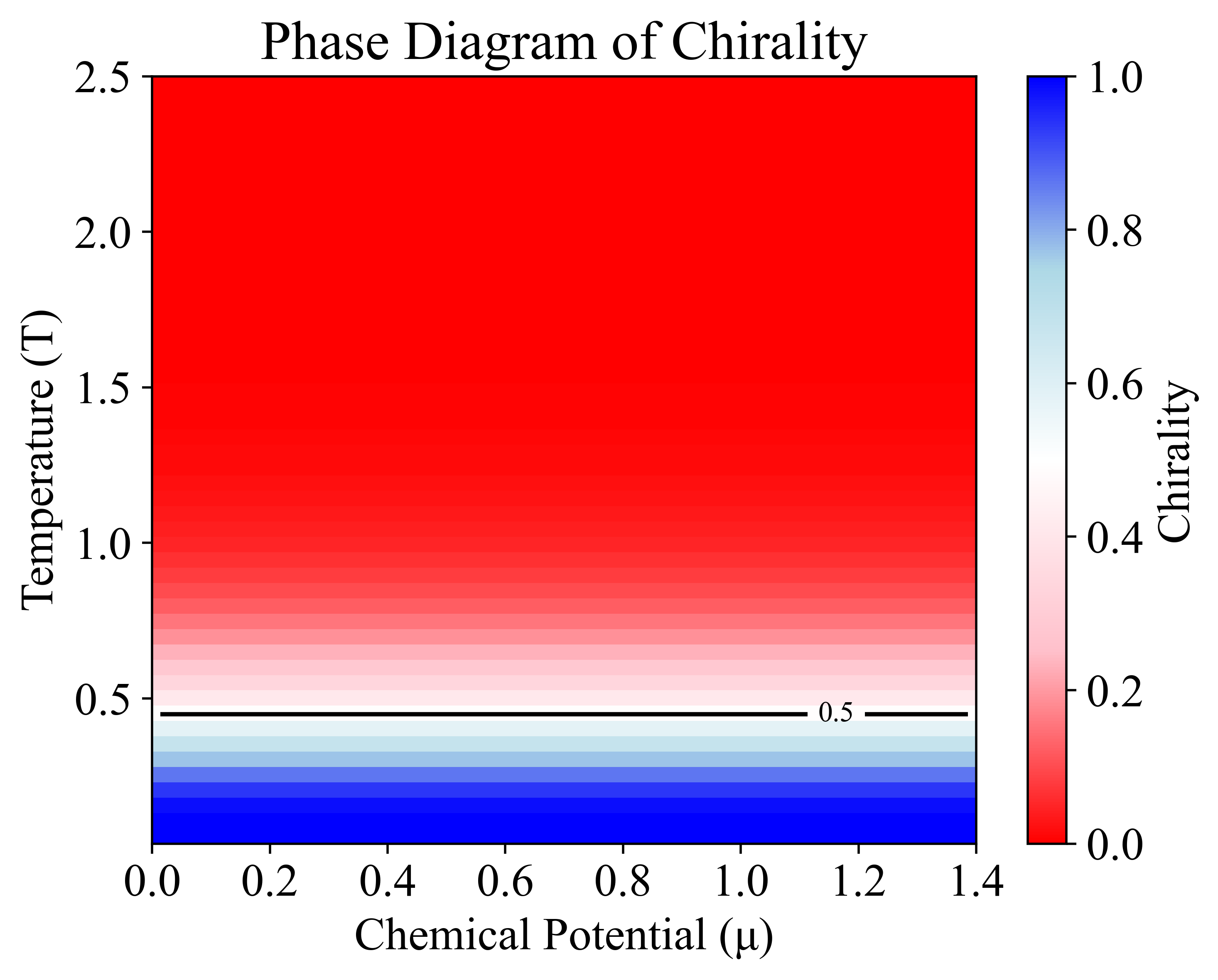}
    \caption{Theoretical chiral condensation result, min-max normalized, independent of $\mu$ because the theory does not take into consideration the chemical potential.}
    \label{theo_chiral}
\end{figure}

The analytical prediction~(\ref{theo_result}) has been plotted in Figure \ref{theo_chiral}. The theory does not consider the influence of chemical potential, so it is not a function of $\mu$. It can be seen that the simulation has given chiral condensation relatively precisely in the low-$\mu$ range.

\section{NN-IQS Model}
The neural network we use in this research is a network with vector nodes inspired by the LIIF model proposed by Chen et al.~\cite{LIIF}. We will transform our training phase diagrams into vectors based on a trainable function $E_{\varphi}$, and then make predictions out of them based on another trainable function $f_{\theta}$,
\begin{equation}
    s = f_{\theta}(z,x)
\end{equation}
where x is target coordinate, z is the nearest vector, and s is the predicted value. The two functions will be trained jointly and both $\varphi$ and $\theta$ will be optimized.

To train the NN-IQS model, we obtain our ground truth training dataset of phase diagrams with $R_{g1}$ $\times$ $R_{g2}$ discrete data points~(it is optional that $R_{g1} = R_{g2}$). We then randomly crop and down-sample them to $R_{i1}$ $\times$ $R_{i2}$ data point ones~($R_{i1} < R_{g1}$ and $R_{g1}/R_{i1} = R_{g2}/R_{i2}$) to prepare the input dataset. Our training task is to predict the $R_{g1}$ $\times$ $R_{g2}$ phase diagram from the $R_{i1}$ $\times$ $R_{i2}$ phase diagram, or in other words, to predict the diagram with more data points from its fewer data point counterpart. Details can be seen in Appendix A and B.

The prediction of a new data point in the middle of existing data points is related to the four nearest existing data points on a rectangular grid. A weighted average will be performed based on relative area $S_{t}$ between the target point and each of the four corner points.
\begin{equation}
    s(x) =  \Sigma_{t \in \{1,2,3,4\}} \, \frac{S_{t}}{S_{1}+S_{2}+S_{3}+S{4}} \, f_{\theta}(z_{t},x-x_{t})
\end{equation}

After training, we will be able to predict the chiral condensation at an arbitrary point on the phase diagram. This means that we have processed the discrete diagrams into continuous ones, and that the whole quantum system is encoded.

Our NN-IQS model offers several notable advantages. First, once the prediction function is trained, it can be applied to all phase diagrams without retraining. This not only reduces computational costs within the dataset but also enables the encoding of new diagrams outside the training set, provided that they share relevant physical characteristics. Second, our training procedure incorporates a special setup in which we can explicitly account for inherent prediction parameter uncertainties. If this setup is properly addressed in advance, training performance can be enhanced. For the phase diagram example, this is achieved by specifying small error intervals along both axes, thereby defining uncertainty bounds for the target coordinates. In practice, this step is automatically included when the up-scaling ratios are specified.

\begin{figure}
    \centering
    \includegraphics[width=1.0\linewidth]{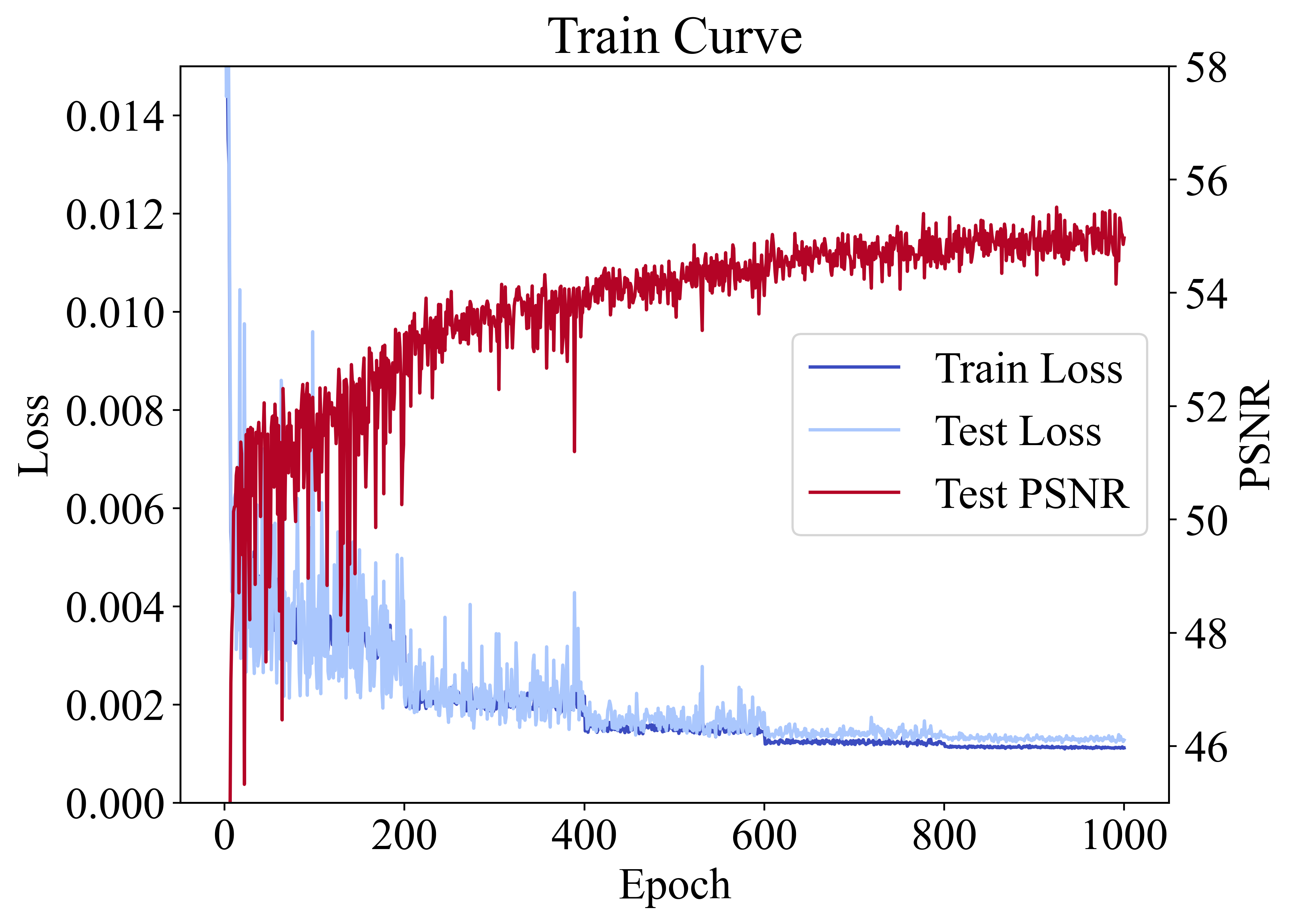}
    \caption{Result of training. Training and test loss are L1 loss, and test PSNR is expressed in dB. Learning rate is gradually decreased.}
    \label{result}
\end{figure}

\section{Results and discussion}
To apply the NN-IQS model to the Schwinger Model, the two axes on the phase diagram would be $T$ and $\mu$. The ground truth training dataset is generated by diagonalizing the Hamiltonian for each specified combination of system size $N$ and $w$~(inversely proportional to lattice parameter a), followed by estimating the chiral condensation on an $R_{g}$ $T$ values $\times$ $R_{g}$ $\mu$ values grid~($R_{g1} = R_{g2} = R_{g}$). The corresponding input diagrams contain $R_{i}$ $\times$ $R_{i}$ data points~($R_{i1} = R_{i2} = R_{i}$). To perform training across different up-scaling ratios from 1 to a maximum of $r_{max}$~($r_{max} \leq R_{g}/R_{i}$), different sub-ground truths are also cropped from the ground truth, details can be seen in Appendix A and B.

Regarding the choice of our $N$ and $w$ values, we need to take into consideration both the validity of the approximation model and computation efficiency. A detailed analysis can be seen in Appendix A.

The complete dataset, consisting of paired inputs and corresponding sub–ground truth diagrams, is randomly divided into training and validation subsets. Model performance during training is quantified using the L1 loss between the predicted and actual chiral condensate values. The validation result is reported in PSNR~(Peak Signal-to-Noise Ratio in dB). In our case where both the predicted and ground truth phase diagrams have $R_{g} \times R_{g}$ data points, the PSNR value is defined as
\begin{align}
     MSE &= \frac{1}{R^{2}_{g}} \, \Sigma^{R_{g}-1}_{i = 0} \Sigma^{R_{g}-1}_{j = 0} \, [P(i,j) - G(i, j)]^{2} \\
    PSNR &= -10 \, log_{10}(MSE)
\end{align}
where $P(i,j)$ and $G(i,j)$ are the predicted and ground truth values at data point $(i,j)$ respectively.

We test the NN-IQS model in four distinct scenarios. First, we consider the basic scenario, where all the statistics are drawn from tests within the training parameter ranges used during training. This serves as a basic guarantee of the model's efficiency. The remaining three scenarios each involve testing a specific parameter outside the training domain. Through these cases, we aim to demonstrate that NN-IQS can: 1) increase data point number on a phase diagram by any arbitrary ratio, even exceeding $r_{max}$; 2) accurately represent phase diagrams corresponding to previously unseen parameter regions; and 3) generalize to represent phase diagrams for previously unseen, larger system sizes.

\subsection{Basic scenario}
First, we examine the model in the most basic post-training scenario, where all test parameters lie strictly within the regions covered during training. During training, we set the ground truth data point number $R_{g} = 196$, and the input data point number $R_{i} = 48$. The maximum up-scaling ratio $r_{max}$ is therefore 4. Since the validation set is drawn directly from the complete original dataset, all validation results in this scenario correspond to parameter regions fully familiar to the model.

The training result is shown in Figure \ref{result}. It can be seen that after a few hundred epochs, the training loss stabilizes at approximately 0.001. Test loss is calculated with the validation set by applying the most recently updated model in each epoch. It is also successfully constrained slightly above the training loss, as expected. The PSNR calculated from validation set, on the other hand, exhibits greater fluctuations, but as training progresses, it is also maintained within $\pm1$dB around a central value of approximately 56 dB. These observations confirm that the NN-IQS model successfully learns the features of the Schwinger model and can accurately predict the chiral condensate after training, without exhibiting signs of overfitting.

We present here a detailed analysis of the relative prediction error across the whole phase diagram with all values converted back to their physical units rather than sigmoid ones, details can be seen in Appendix A). Specifically, each predicted diagram is compared to its ground-truth counterpart through an element-wise absolute-value subtraction, followed by element-wise division by the ground-truth values. This procedure yields a new map that specifies the relative error at every data point in the phase diagram. This operation is applied to every diagram in the validation set for all up-scaling ratios considered, and the resulting data are compiled for subsequent analysis.

\begin{figure}
    \centering
    \subfigure[]{ \includegraphics[width=1.0\linewidth]{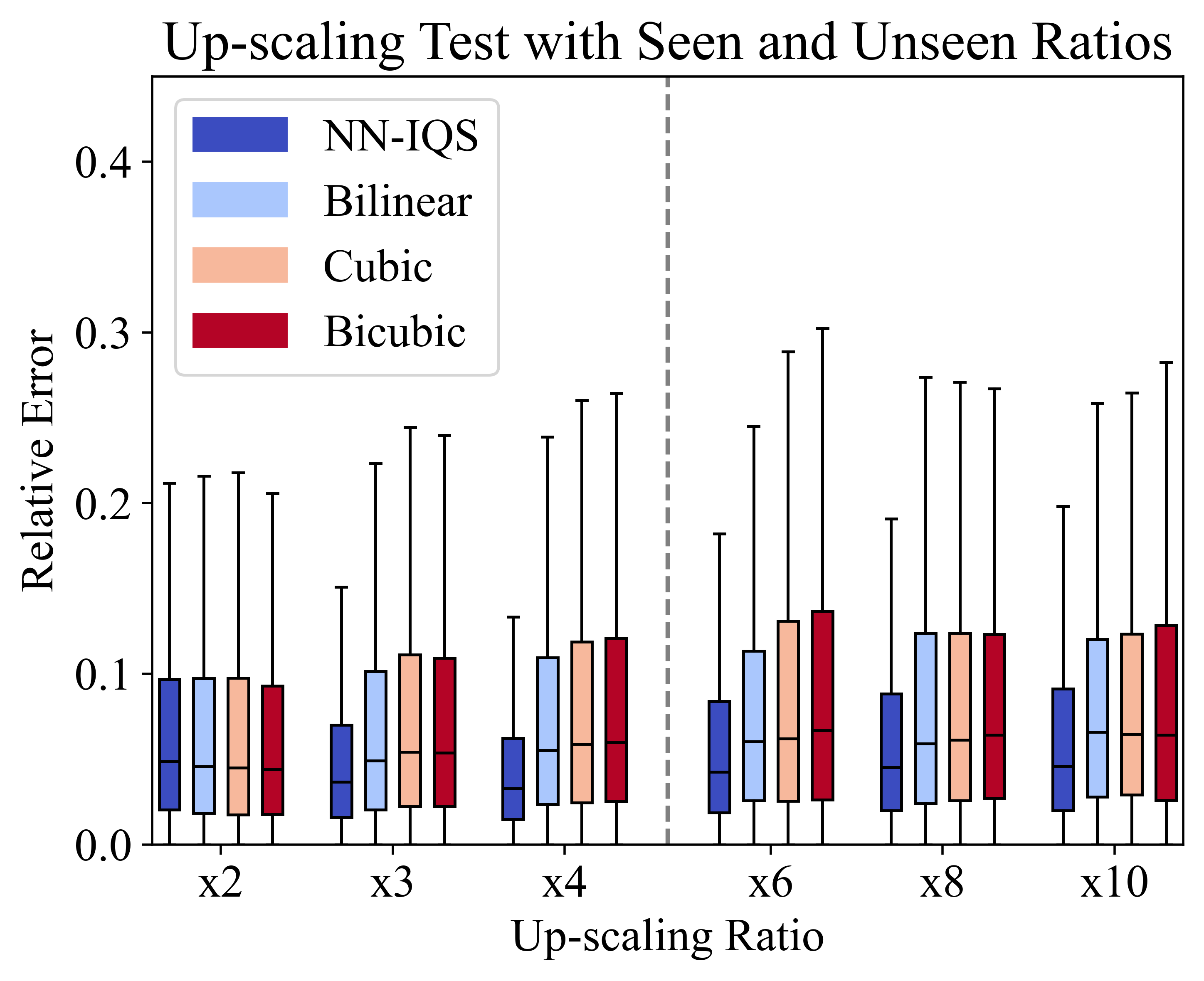}}
    \subfigure[]{ \includegraphics[width=1.0\linewidth]{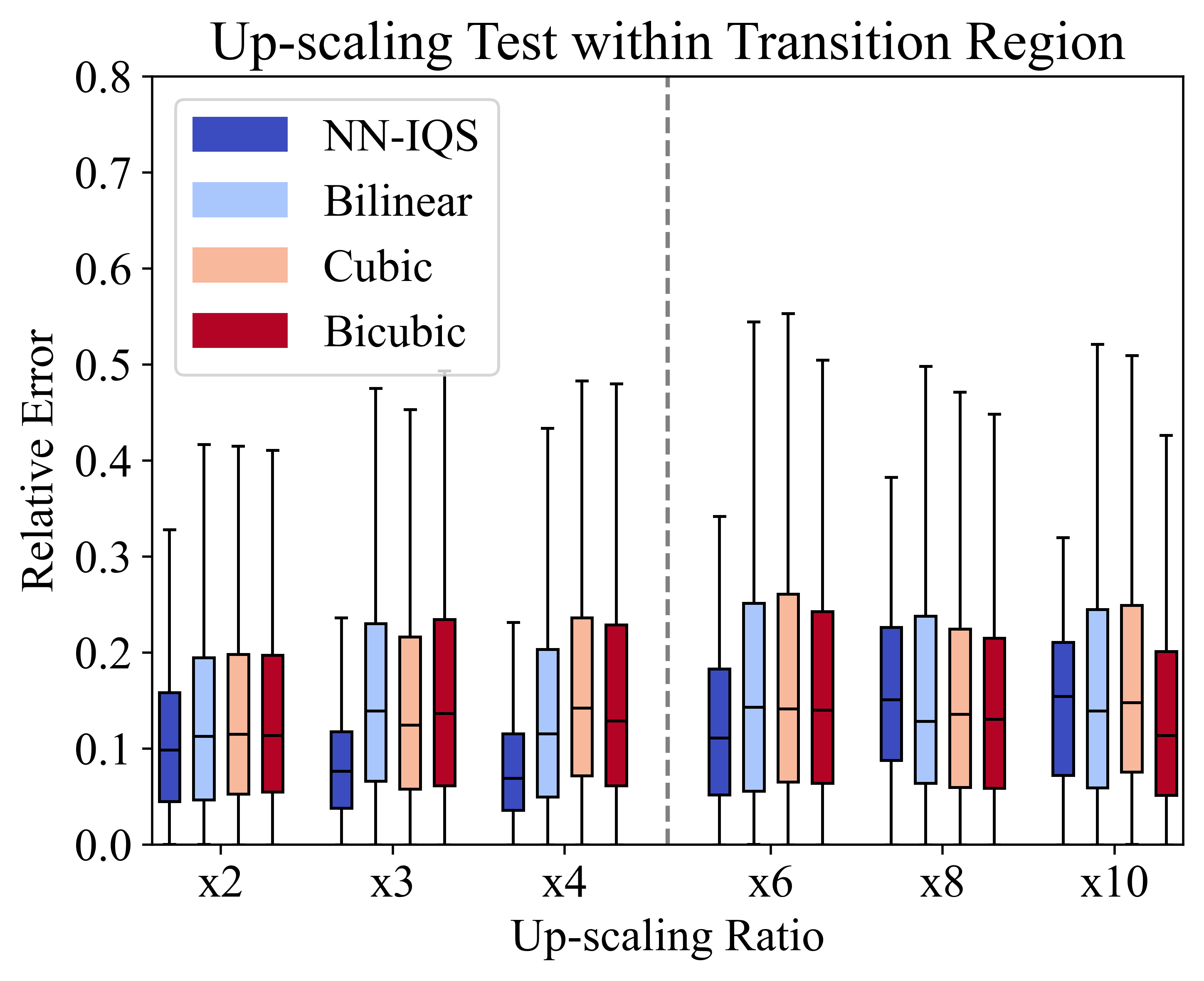}}
    \caption{Relative error at different up-scaling ratios within (a) the whole phase diagram, (b) transition region. Three interpolation methods are plotted as benchmarks. $Y$ axis scales differ between the two figures for a clear display, it is expected that error will generally be larger within transition region.}
    \label{in_and_out_diff}
\end{figure}

We present the results of $\times$2, $\times$3, and $\times$4 here. It can be seen in the left half of Figure \ref{in_and_out_diff}(a) that, the mean relative error remains at or below 5\% in $\times$3 and $\times$4 up-scaling ratios. This indicates, in general, a high degree of confidence in the NN-IQS model’s ability to deliver high-precision predictions across the entire phase diagram.

The observation that the lowest ratio, $\times$2 up-scaling, yields the weakest performance may appear counterintuitive, and we suggest an explanation here. In our test, random cropping and down-sampling process are applied between our general ground truth, sub-ground truth and input, details can be seen in Appendix A. These processes guarantee randomness and generality, but can also introduce certain issues. Specifically, at low up-scaling ratios, it is very likely that the intersection area between our input and corresponding sub-ground truth happens to be lower than expected. For example, if the selected sub–ground truth data points are widely distributed across the spatial domain but the corresponding input points are concentrated predominantly in one region of it, the remaining region of the sub–ground truth becomes effectively unpredictable. At higher up-scaling ratios, the probability of this mismatch decreases, mitigating the effect.

For benchmarking purposes, we also present results obtained from purely interpolated phase diagrams. These represent the most straightforward approaches to generate the phase diagram with more data points from its fewer counterpart. It is worth noting that for a 2D diagram, no universally accepted interpolation strategy exists, thus different methods can be seen in literature. Here, we select three widely used techniques: bilinear interpolation, separate coordinate cubic interpolation, and simultaneous bicubic interpolation. These methods encompass a large portion of practical interpolation implementations in modern engineering. It can be seen in Figure \ref{in_and_out_diff}(a) that, the NN-IQS model consistently and substantially outperforms all interpolation strategies tested.

It is important to note that the relative error computed over the entire phase diagram does not necessarily reflect the true predictive power of the model. This is because the regions of greatest physical interest are typically the transition regions, whereas large portions of the diagram correspond to extreme~(maximum or minimum) values where interpolation methods often perform trivially well. To address this, we also report statistics collected only within the transition region in Figure \ref{in_and_out_diff}(b). From this figure, it is evident that the NN-IQS model consistently outperforms all interpolation benchmarks for all metrics within the trained up-scaling ratios. In $\times$3 and $\times$4 tests, the trained model achieves more than twice the precision of the benchmarks. In unfamiliar up-scaling ratios, the benchmarks may occasionally yield lower mean errors. However, the NN-IQS model maintains the maximum error within a relatively small and acceptable range, indicating greater overall reliability.

\begin{figure*}
    \centering   
    \subfigure[]{ \includegraphics[width=0.4\linewidth]{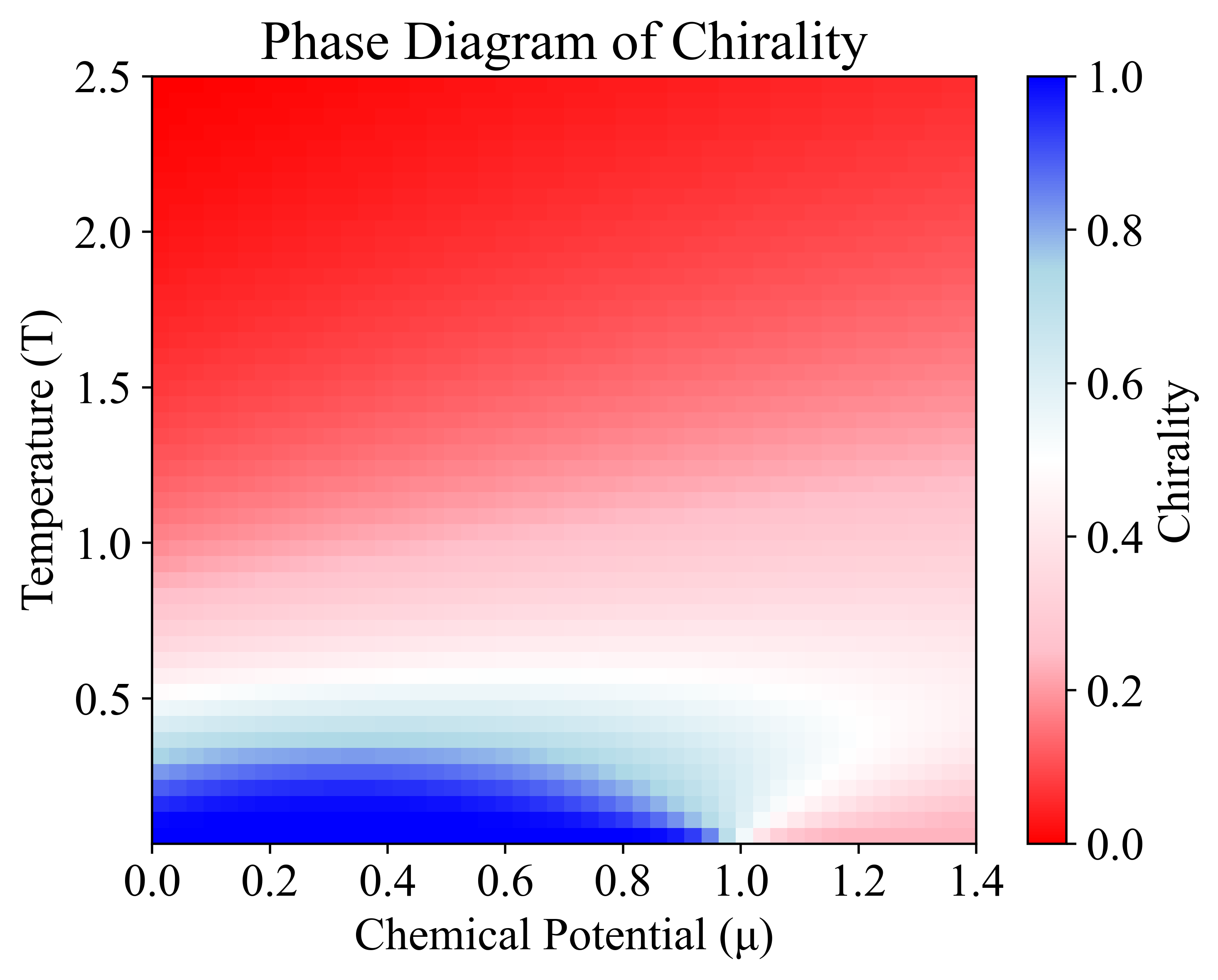}}%
    \subfigure[]{ \includegraphics[width=0.4\linewidth]{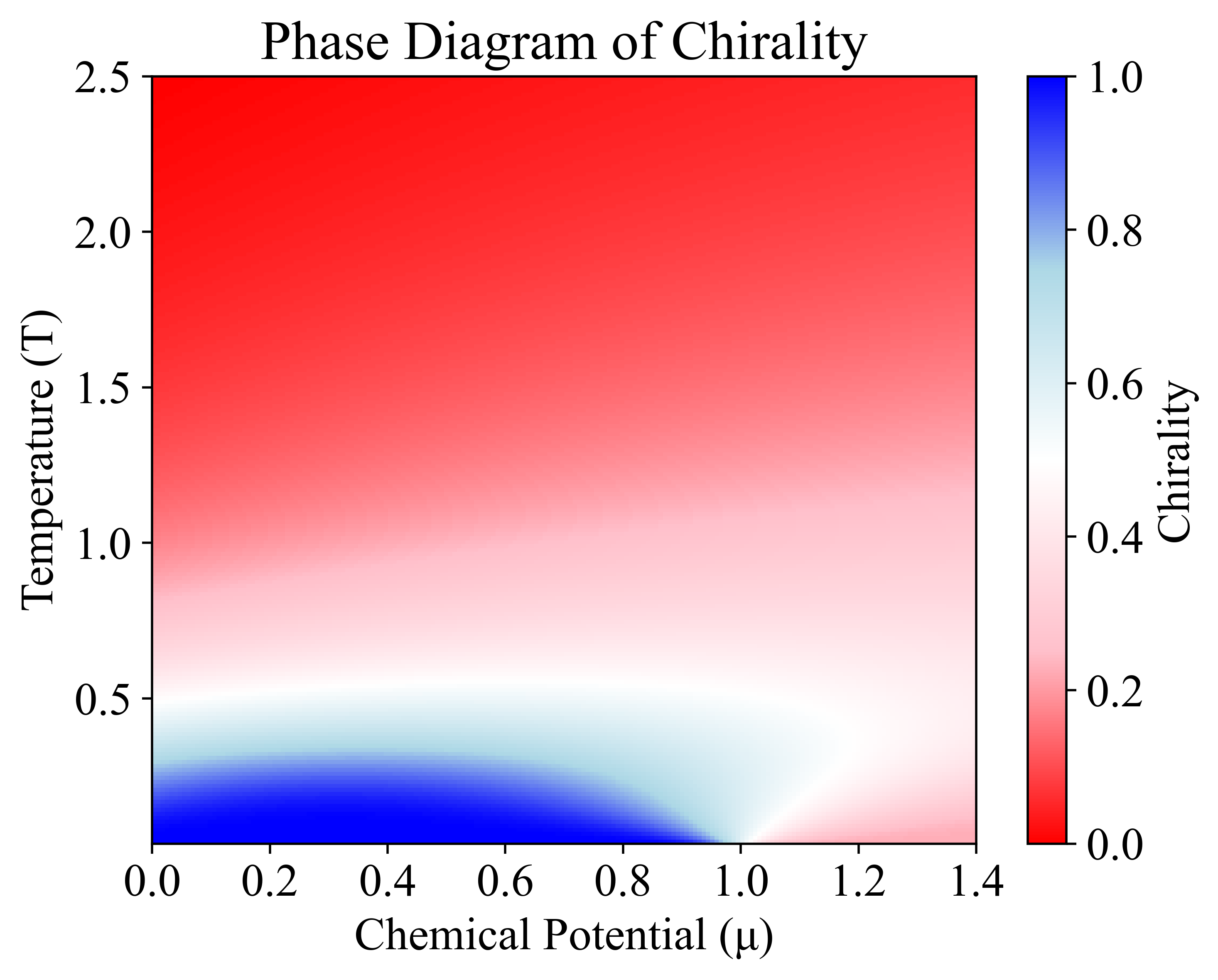}}
    \subfigure[]{ \includegraphics[width=0.4\linewidth]{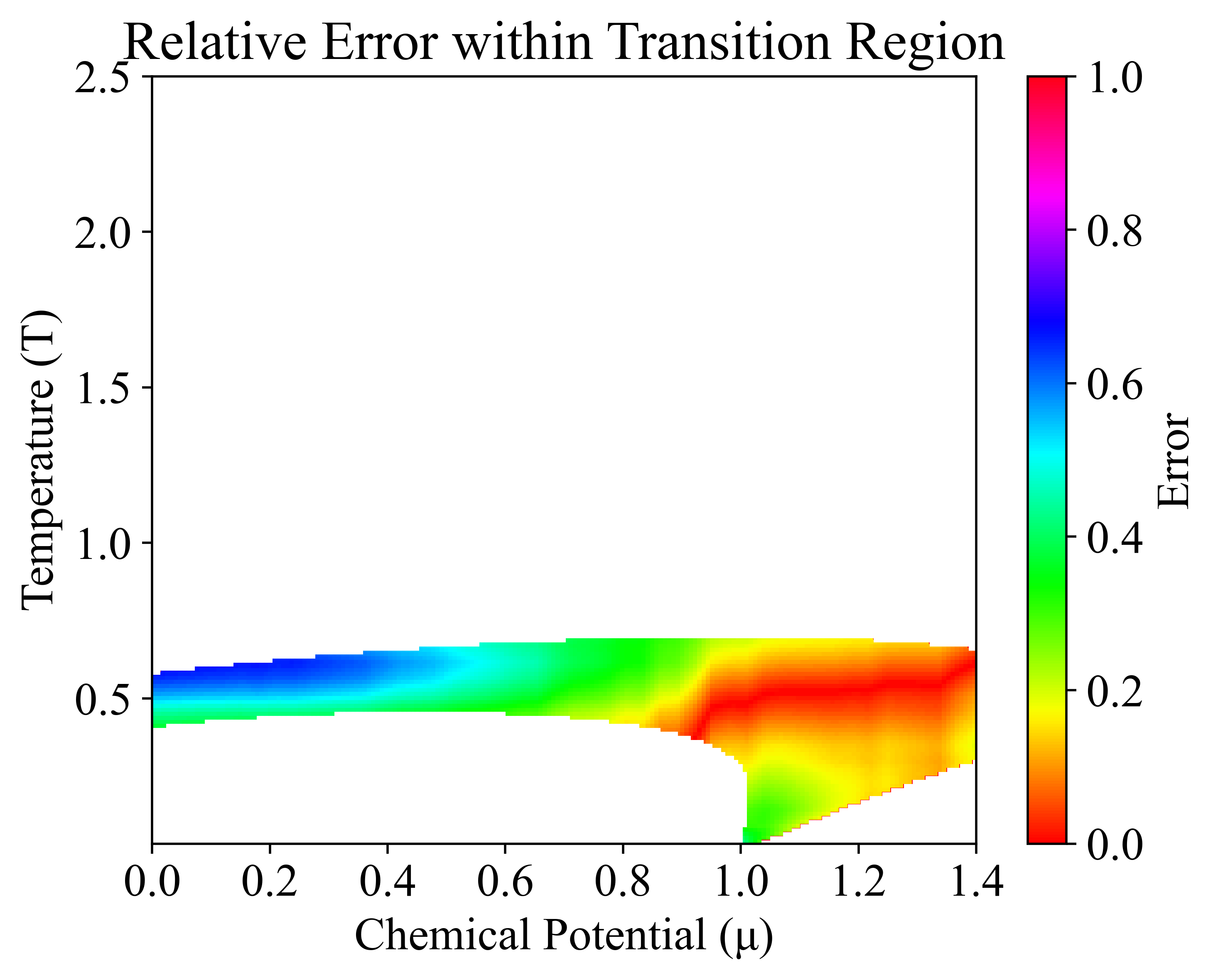}}%
    \subfigure[]{ \includegraphics[width=0.4\linewidth]{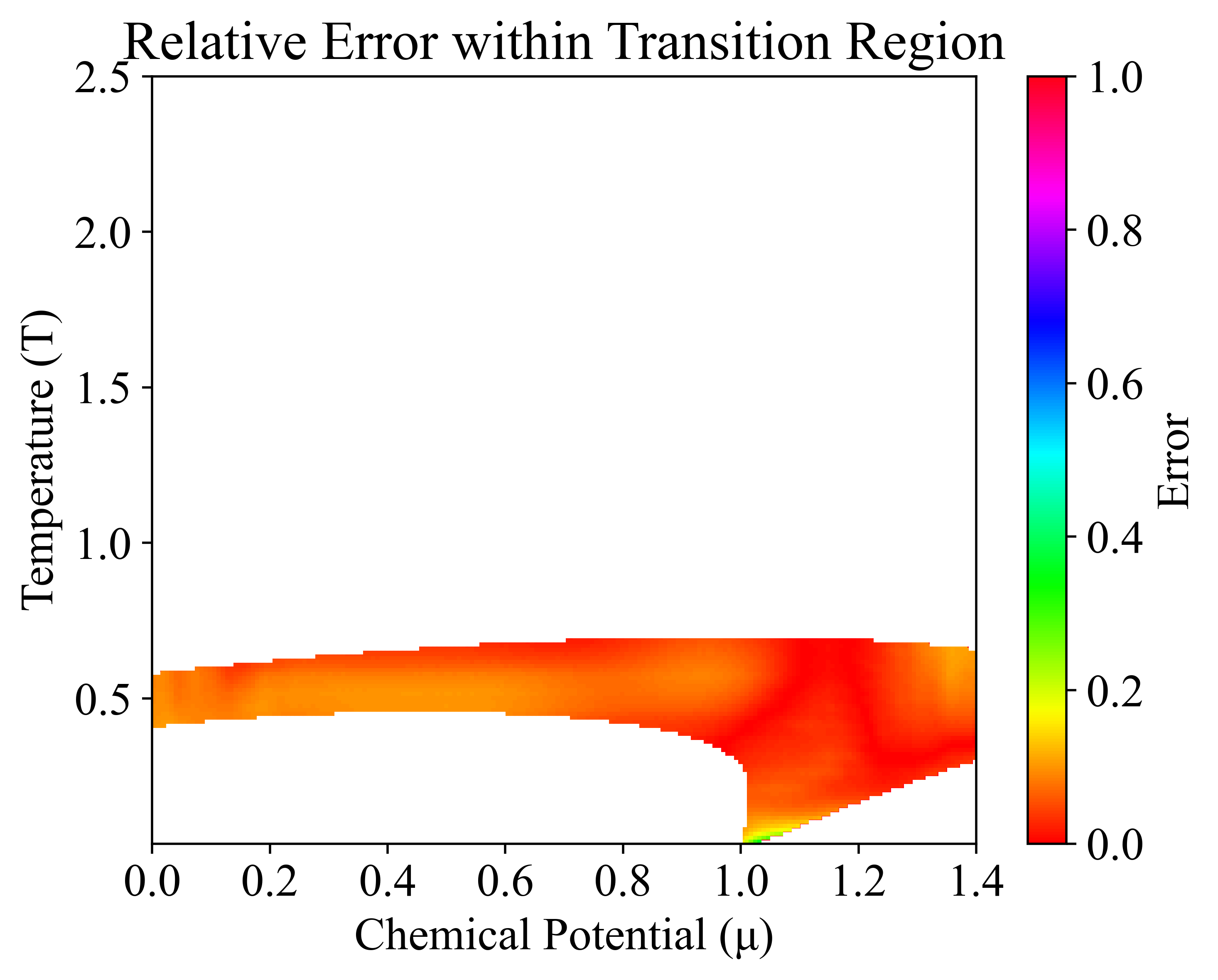}}
    \caption{(a) Simulated phase diagram with 48$\times$48 data points. (b) Simulated phase diagram with 192$\times$192 data points. (c) Relative error within transition region from bilinear interpolation prediction. (d) Relative error within transition region from NN-IQS prediction. Transition region is calculated from (b).}
    \label{upscale-test}
\end{figure*}

To further illustrate this gap in precision, we present here a realistic example of prediction. As shown in Figure \ref{upscale-test}, we select a typical sample point out of the original dataset but within the parameter range, with $N = 8$ and $w/g = 1$. We then generate two phase diagrams out of it, one with 48$\times$48 data points, as shown in Figure \ref{upscale-test}(a), another with 192$\times$192 data points, as shown in Figure \ref{upscale-test}(b). The objective is to predict (b) out of (a) with minimum error. That is, to do a $\times$4 up-scaling with standard input. The relative error in transition region for bilinearly interpolated and NN-IQS predicted results are presented in Figure \ref{upscale-test}(c) and (d) respectively. It is apparent that the NN-IQS model maintains the relative error across the entire region at a very low level, whereas the interpolation benchmark fails to control the error, yielding unreliable performance. This example underscores that the NN-IQS model provides a robust and trustworthy approach to up-scaling phase diagrams.

To assess the extensibility of the model, we further examine its performance on test cases involving up-scaling ratios, parameters, and system sizes not encountered during training.

\subsection{Beyond up-scaling ratios}
Beyond the basic scenario, we also investigate whether the NN-IQS model can up-scale phase diagrams to ratios that were not included in its training. For this purpose, we generate an additional test dataset consisting of phase diagrams with substantially more data points for test use. Previously in the basic datasets, each of our phase diagrams has 196$\times$196, or 38416 data points. In this test however, we generate a new dataset with phase diagrams containing 480$\times$480, or 230400 data points.

With this new dataset, we evaluate the capability of the NN-IQS model to upscale phase diagrams by factors ranging from $\times$6 to $\times$10, which are beyond $r_{max}$ and are therefore entirely unfamiliar to the model from its training stage. To execute a fair test, we ensure the input diagrams~(after random down-sampling) still at least have a data point number of 48$\times$48, consistent with training. The same relative error analysis described previously is applied to this new dataset, and we present the results of $\times$6, $\times$8, and $\times$10 here.

As shown in the right half of Figure \ref{in_and_out_diff}(a) and (b), although the NN-IQS model has never been trained to up-scale a diagram by more than a factor of five, it maintains consistent performance at higher ratios. The accuracy is only marginally lower than that achieved within the trained ratio range, with the relative error remaining below 7\% across the entire diagram.

\subsection{Uncovered lattice parameters}
We also evaluate the model’s ability to generalize to unfamiliar $w$ regions. In this case, no new datasets are generated, as the $w$ values used in the basic scenario training already span nearly the entire physically reasonable range. Instead, we partition the initial $w$ range into a central interval and two side intervals. A new round of training is then performed on data from one of these ranges, with the goal of predicting phase diagrams in the other ranges~(see Appendix A for details). As in previous tests, we report the relative error distributions for both the entire diagram and the transition region.

\begin{figure}
    \centering
    \includegraphics[width=1.0\linewidth]{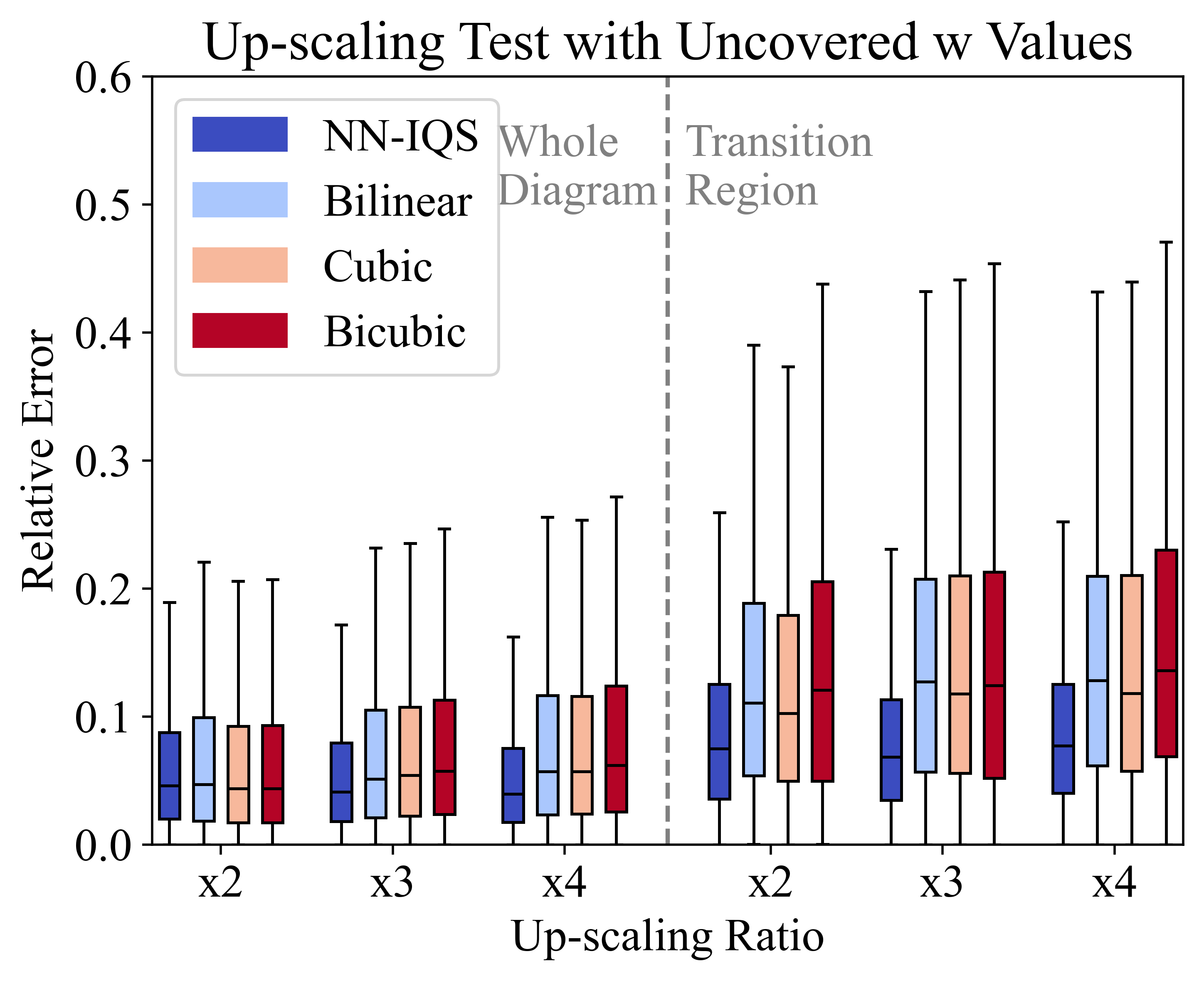}
    \caption{Relative error at different up-scaling ratios with uncovered w values. $\times$2, $\times$3, and $\times$4 results are presented. Error within the whole diagram and transition region only are presented on the two sides in the same $Y$ axis scale. As expected, predictions in the transition region are generally more challenging and yield larger errors.}
    \label{unseen_w}
\end{figure}

Given the limitation of available computational resources, we only present here~(with our initial dataset) the result of $\times$2 to $\times$4 with uncovered $w$ values in Figure \ref{unseen_w}. By using same scale on the y axis, we emphasize once again that the relative error is generally larger within the transition region than over the entire phase diagram, in agreement with physical intuition.

It can be seen clearly that although we are testing with unseen parameters, as long as the parameter is still in the reasonable range, our NN-IQS model shows an apparent advantage over all interpolation results. This demonstrates the model’s potential to generalize effectively to unobserved parameter domains. In particular research interest, in some physical problems, there may be some parameter regions inaccessible to either theoretical analysis or experimental measurement. Our NN-IQS model can then be used to provide significant insight into the problem from a numerical perspective. 

\subsection{Out of reach system size}
Another promising extrapolation possibility we like to propose here, is the extension to larger system sizes. A widely recognized challenge in modern quantum algorithms and experimental platforms is the scaling to systems with more qubits. This difficulty is typically accompanied by exponentially increasing complexity in qubit connectivity and interactions, making the design of efficient algorithms highly nontrivial. If simulations could be extended to larger systems without requiring upgrades to the underlying physical hardware, a wide range of large-scale quantum simulations would become feasible in the NISQ era. For this reason, we present here the results demonstrating the ability of our model to generalize to larger system sizes.

As the system size increases, exact diagonalization requires exponentially longer computation time. While this does not affect the training or application of our model, it poses a challenge for evaluation, as obtaining the ground-truth data for larger systems becomes computationally demanding. In this test, we therefore simulate only a single phase diagram with $N = 12$~(which is out of the training range) while keeping all other settings identical to the corresponding sample point in the basic scenario. As before, we focus on the relative error within the transition region. It should be noted that, owing to the change in system size, the transition region in this larger diagram differs from that in the smaller system.

\begin{figure*}
    \centering   
    \subfigure[]{ \includegraphics[width=0.4\linewidth]{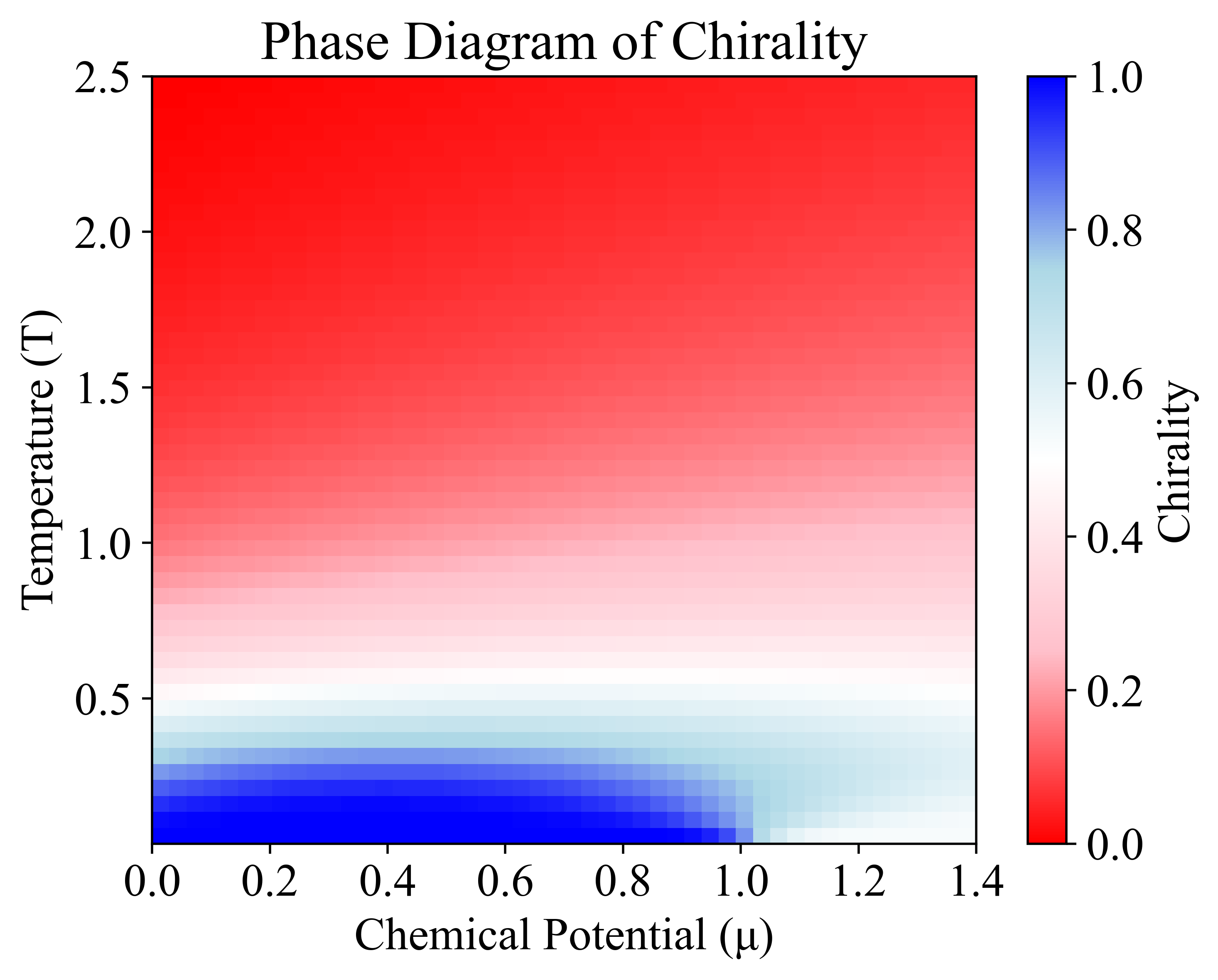}}%
    \subfigure[]{ \includegraphics[width=0.4\linewidth]{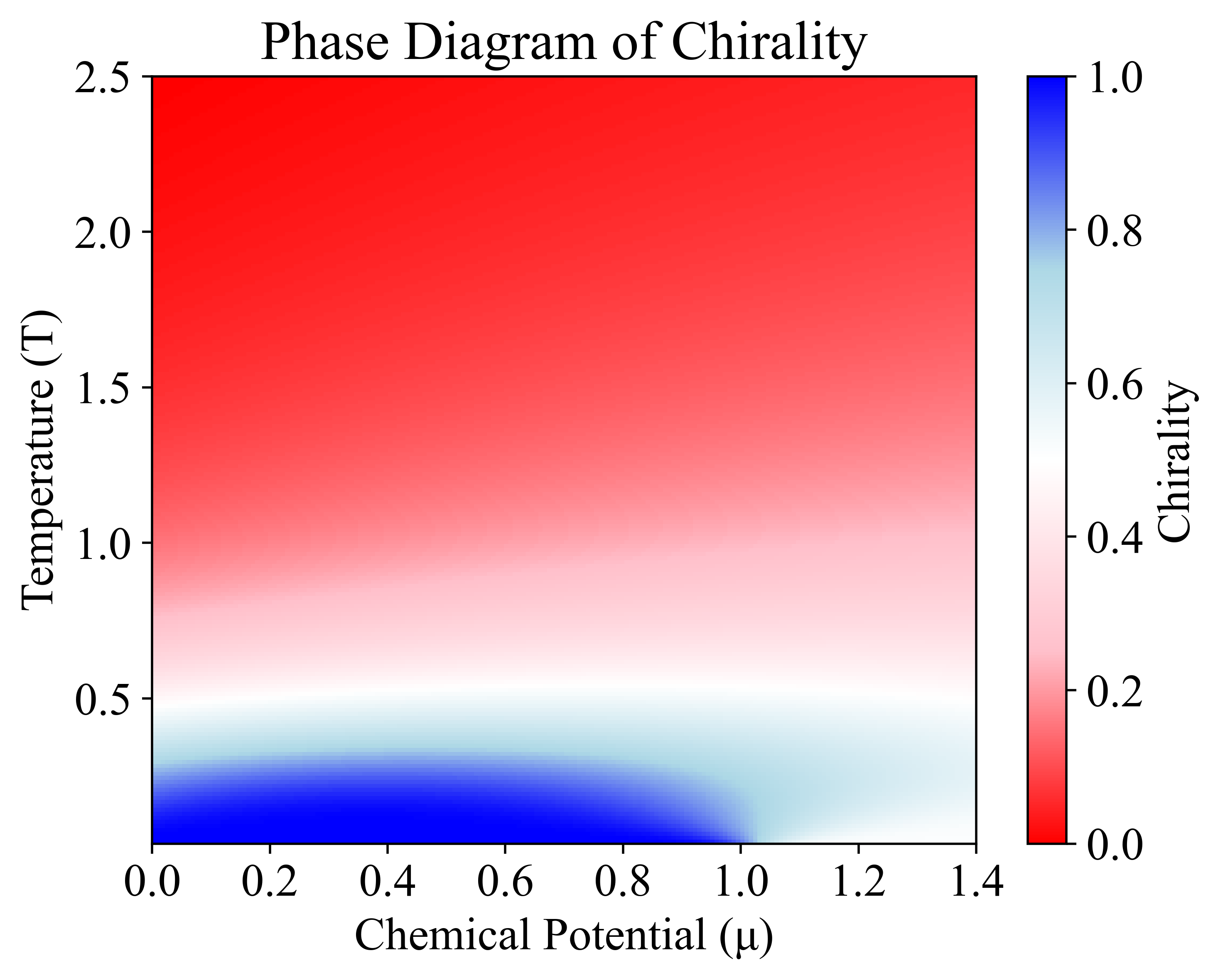}}
    \subfigure[]{ \includegraphics[width=0.4\linewidth]{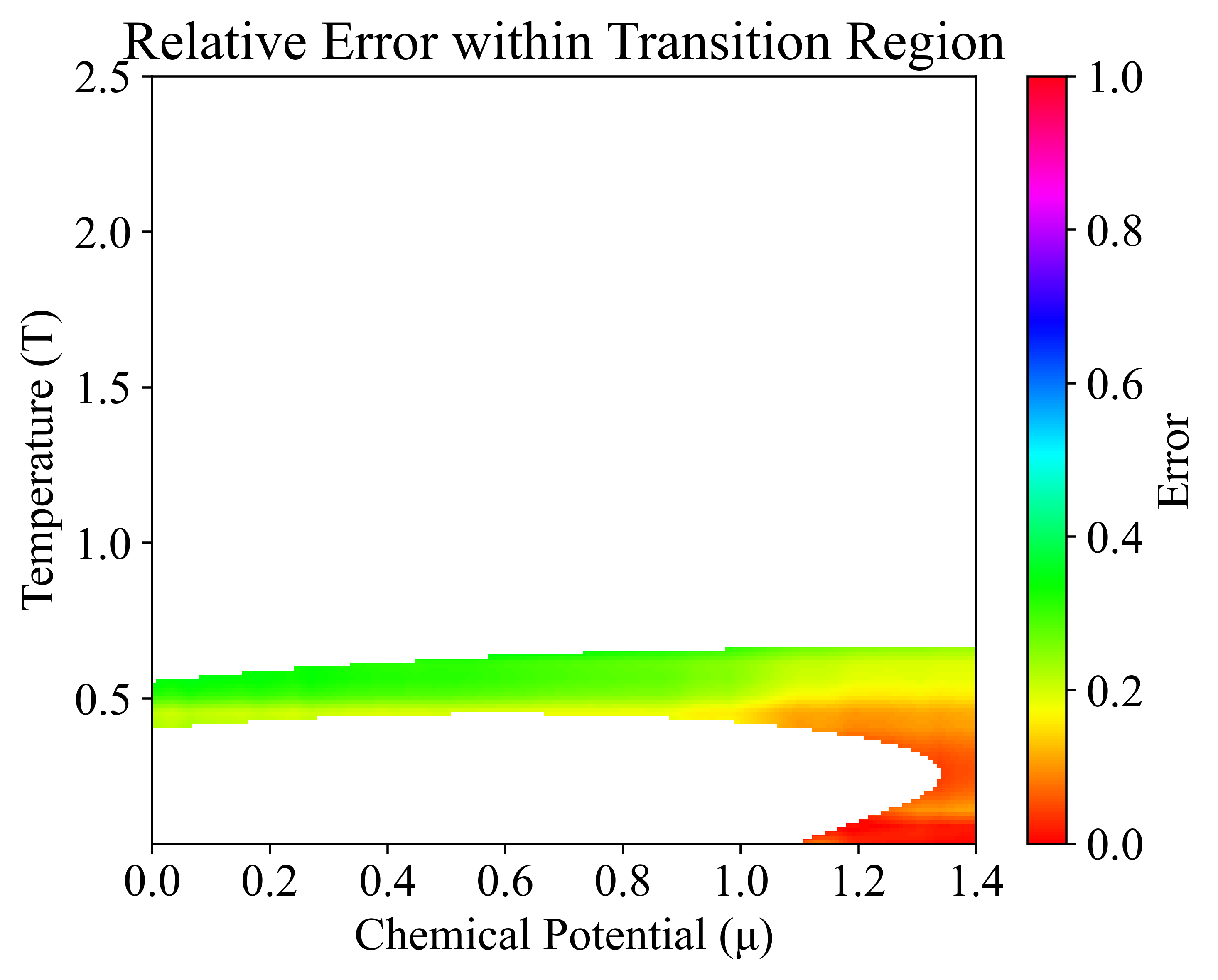}}%
    \subfigure[]{ \includegraphics[width=0.4\linewidth]{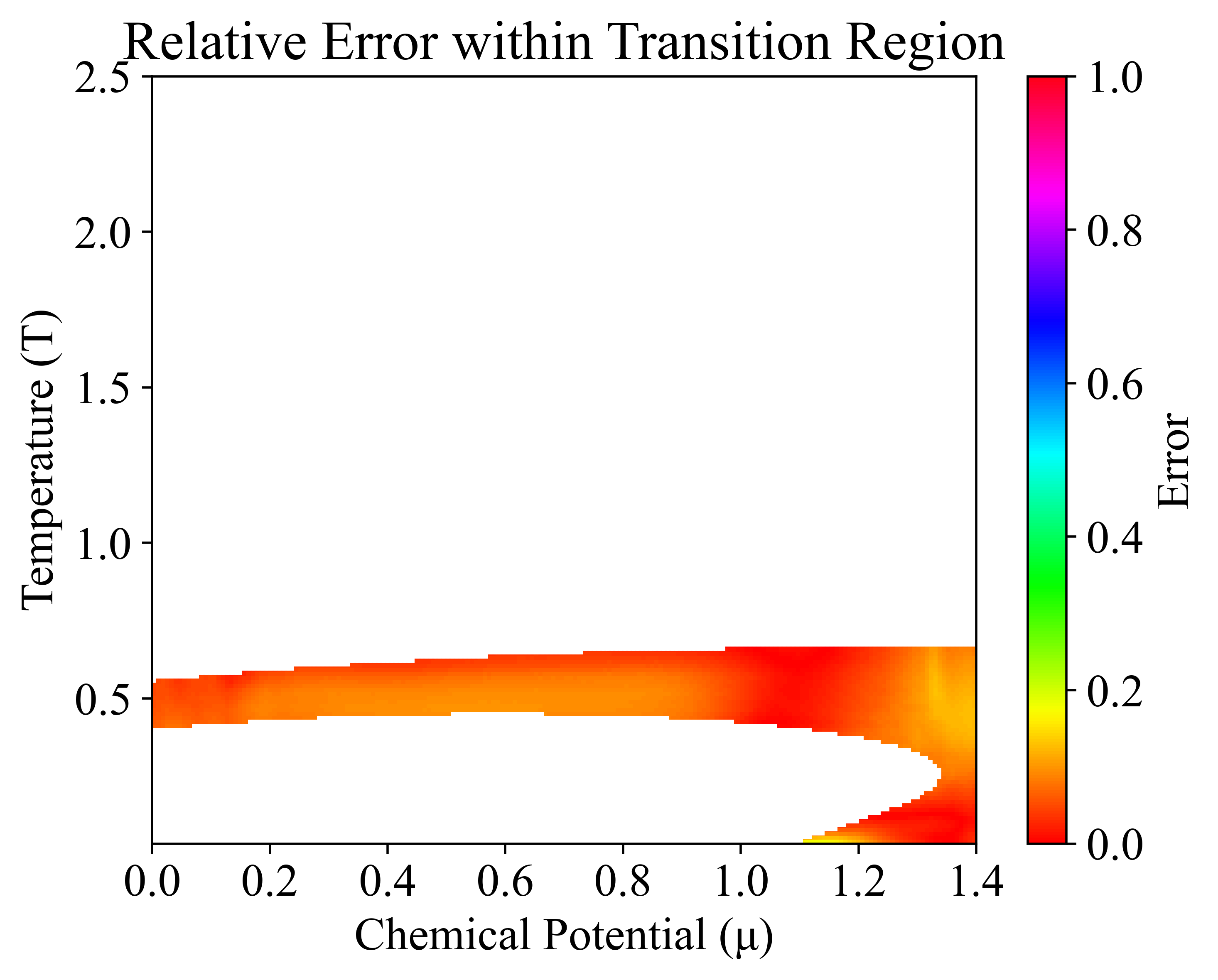}}
    \caption{(a) Simulated phase diagram with 48$\times$48 data points. (b) Simulated phase diagram with 192$\times$192 data points. (c) Relative error within transition region from bilinear interpolation prediction. (d) Relative error within transition region from NN-IQS prediction. Transition region is calculated from (b). System size is 12 here, and the transition region is different from that in Figure \ref{upscale-test}.}
    \label{N12}
\end{figure*}

It can be seen in Figure \ref{N12} that, although $N = 12$ lies entirely outside the training range and exhibits a transition region distinct from that of the smaller systems, the NN-IQS model still achieves high accuracy within this new transition region, outperforming Bilinear interpolation benchmark. We anticipate similar performance for even larger systems, given that as long as they share similar parameters, they would exhibit similar physical properties, which our NN-IQS model can capture easily. In this way, no matter what system size is readily available from simulation, our NN-IQS model can always push the boundary further. This result suggests a promising route for exploring large-scale quantum problems that would otherwise remain inaccessible to direct quantum simulation alone.

To be precise, we briefly note that different formulations of the Schwinger model can give rise to distinct physical phenomena. In particular, when appropriate topological parameter and fermion mass are included, the model can exhibit explicit first-order transition lines and second-order critical points~\cite{transition1, transition2}, bringing it into closer correspondence with more complex gauge theories like QCD. However, numerical simulations in these regimes become exponentially costly near the transition regions owing to severe sign problem~\cite{transition2, signprob1}. Moreover, existing results for two-dimensional phase diagrams of such systems are scarce. We therefore defer a systematic investigation of this sector to future work.

\section{Conclusion and outlook}
In this work, we introduce a new model which we call the NN-IQS model. We use a neural network to make an implicit representation of a quantum system, the Schwinger Model. 

We first evaluate the NN-IQS model in a basic scenario, where all parameters lie within the range used during training. In this setting, the predicted results clearly surpass the benchmark, particularly in the transition region where the chiral condensate changes rapidly and interpolation methods break down. We then conduct further tests on entirely unfamiliar parameter regimes, including up-scaling ratios beyond the training range, previously uncovered lattice parameters, and a larger system size. In all cases, the NN-IQS model consistently outperforms the benchmark methods.

These results demonstrate that the NN-IQS model remains reliable for massive and rapid sampling, even in challenging scenarios such as phase transition regions, previously inaccessible parameter regimes, and larger system sizes. Consequently, it offers the potential to enable precise numerical investigations of quantum systems that face these obstacles, making it a powerful tool for advancing studies in the NISQ era.

Due to the computational resources currently available, our study is limited to simulators, with results reported for up-scaling ratios up to 10 and system sizes up to 12. Despite that, the workflow is compatible with current quantum platforms. A detailed discussion of experimental feasibility, including thermal-state preparation protocols and candidate hardware architectures, is provided in Appendix~\ref{app:feasibility_of_physical_implementation}. Meanwhile, apart from the Schwinger Model, we further anticipate that the NN-IQS framework can be extended to other quantum systems, most notably the QCD model, where it could provide valuable insights for both theoretical and experimental investigations. These directions are left for future work.

\section{Acknowledgments}
The authors are grateful for the valuable discussions with Dr. Yan Zhu and Prof. Yingying Li. This work is supported by the Quantum Science Center of Guangdong–Hong Kong–Macau Greater Bay Area, the NSFC/RGC JRS grant~(RGC Grant No. N$\_$HKU774/21), the GRF of Hong Kong~(Grants No. 17310622 and No. 17303023), and Guangdong Provincial Quantum Science Strategic Initiative (Grant No. GDZX2404001).

\appendix

\section{Detailed set-ups in simulation, learning, and test}
All theoretical calculations in this study, including the ground-truth phase diagrams for all datasets and examples, are performed via quantum simulation using QuTiP and NumPy. The simulation operators are constructed from the spin model described above, with Kronecker products employed to implement parallel operations across separate qubits. For finite-temperature cases, the Gibbs state is first prepared, after which exact diagonalization is applied to obtain the expected value of the chiral condensate.

In specifying the Schwinger model parameters used in this work, we treat $N$ and $w$ values separately. Since the Schwinger model serves as a simplified analogue of QCD, the choice of system size $N$ has a significant impact on the accuracy of this approximation. $N$ values below 6 are generally regarded as oversimplified~\cite{betavqe}, and that above 10 are tested computationally inefficient. Thus, we generate our dataset with different N values 6, 8, and 10. As for the $w$ values, or the ratio $w/g$ that we apply, it can be seen in the model Hamiltonian that $w/g$ must be restricted within a particular region in order for the Hamiltonian to make sense~(the Hamiltonian will explode either when $w/g$ approaches zero or infinity). In~\cite{betavqe}, a range of [0.35, 0.5] was applied for $w/g$ in the dataset, while in our case, to test the robustness of the NN-IQS model across a wider variety of scenarios, we adopt a significantly broader range of [0.3, 1.5] for the training set.

For every phase diagram generated, expected value is calculated separately at every data point. Along the $\mu$ axis, data points start at $\mu = 0$, while along the $T$ axis, they start from $T = 0.1$ to avoid possible singularity issues at $T = 0$. The upper limits are set to $\mu/g=1.4$ and $T/g=2.5$. Before being fed into the neural network, whether in the training, validation, or testing stages, a sigmoid function is applied to all data points to improve convergence during optimization. The final outputs in all demonstrations are then transformed back into their physical values to illustrate real-world applicability.

During training of the NN-IQS model, the training epoch is set to 1000. Initial learning rate is chosen to be 0.00001, which will gradually decrease at epoch 200, 400, 600, and 800.

In both the training and testing stages, we define sub-ground truths for each up-scaling ratio. For an up-scaling ratio of $n$, the sub-ground truths have a data point number of $48n\times48n$ because we keep the input data point number to be 48$\times$48. The sub-ground truths are cropped from the general ground truth in the datasets~(either with $196 \times 196$ data point number or $480 \times 480$ data point number) in two different ways. During training, we follow the cropping strategy of the original LIIF study, extracting a contiguous square block of the desired size directly from the original diagram. Whereas in the test stage, the required number of coordinate values are randomly sampled along both the $\mu$ and $T$ axes to assemble the target phase diagram. A graphical illustration of this process is provided in Appendix B. After cropping, the sub-ground truths are randomly down-sampled to 48$\times$48 data points counterpart to construct the inputs. The training task would then be to recover the cropped sub-ground truths from the inputs.

In both the training and testing stages, the cell values are naturally defined to be inversely proportional to the number of output data points along a single axis, reflecting the intuition that the more the output data points, the more precise the coordinates and the smaller the physical uncertainty.

To evaluate the relative error during testing, the predicted phase diagrams are compared with the corresponding ground truths on an element‑wise basis. Both predictions and ground truths are first transformed back from their sigmoid‑scaled values, after which the absolute relative error is computed for each data point pair. The resulting output is a complete error map, covering either the entire diagram or, in certain cases, only the transition region, showing the local relative error at each point. For all box plots, data points classified as statistical outliers are omitted, following the standard criterion that any value lying more than 1.5 times~($Q_{3}-Q_{1}$) beyond the interquartile range is considered an outlier.

For testing on uncovered lattice parameters, we perform an additional training procedure in which the initial dataset is split into two disjoint parts. The new training and validation sets are constructed from phase diagrams with $w/g$ values within $[0.5, 1.3]$. The unseen test set consists of the remaining phase diagrams, corresponding to $w/g$ values within $[0.3, 0.499]$ and $[1.301, 1.5]$.

\section{Detailed description of different cropping methods used in dataset preparation}
\begin{figure}
    \centering
    \includegraphics[width=0.9\linewidth]{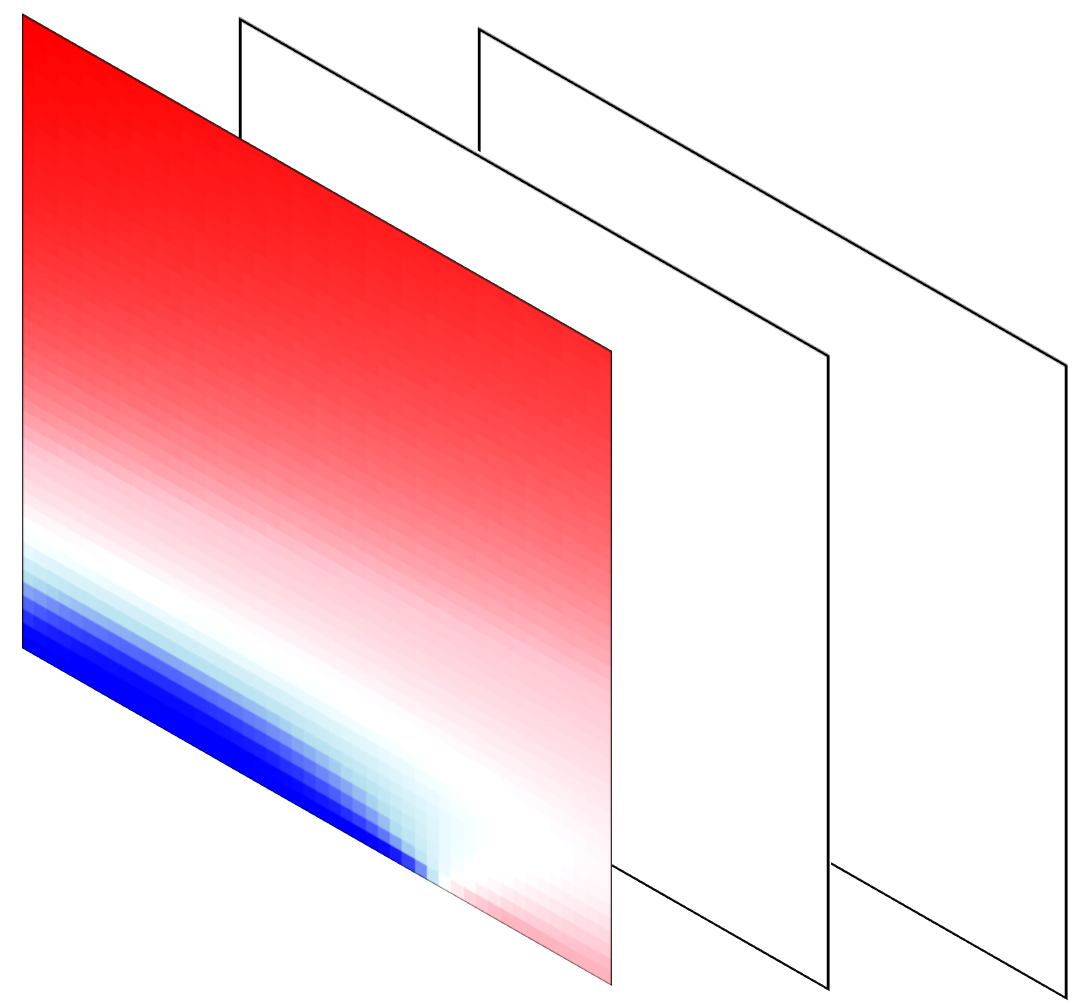}
    \caption{Phase diagrams are represented as 3D arrays, with one channel containing the physical data and the remaining two channels filled with uniformly low values. In the visualizations, these two inactive channels are shown in white to indicate their current placeholder role; they remain available for incorporating additional data in future studies.}
    \label{3D}
\end{figure}

The initial concept for the NN-IQS model is inspired by classical image-processing algorithms applied to color images, where a neural network must simultaneously process all three color channels~(RGB) to produce a comprehensive prediction. In our NN-IQS model however, we consider only one layer of values, namely the chiral condensation values on our 2D phase diagrams. Therefore, the algorithm is designed to transform the phase diagrams to 3D arrays before input to resemble the classical idea and enable more flexibility for further studies~(maybe with more than one layer of quantum data), as shown in Figure \ref{3D}.

The three channels in the input structure can in principle each store a separate 2D array. In our present implementation, one channel contains the physical information from the phase diagrams. The remaining two channels are filled with uniformly small constant values to ensure that they neither interfere with the target data layer nor introduce numerical overflow issues during computation. After prediction, the 3D output is separated back into its constituent layers, and the relevant channel is extracted to recover the original 2D phase diagram for analysis.

\begin{figure*}
    \centering   
    \subfigure[]{ \includegraphics[width=0.8\linewidth]{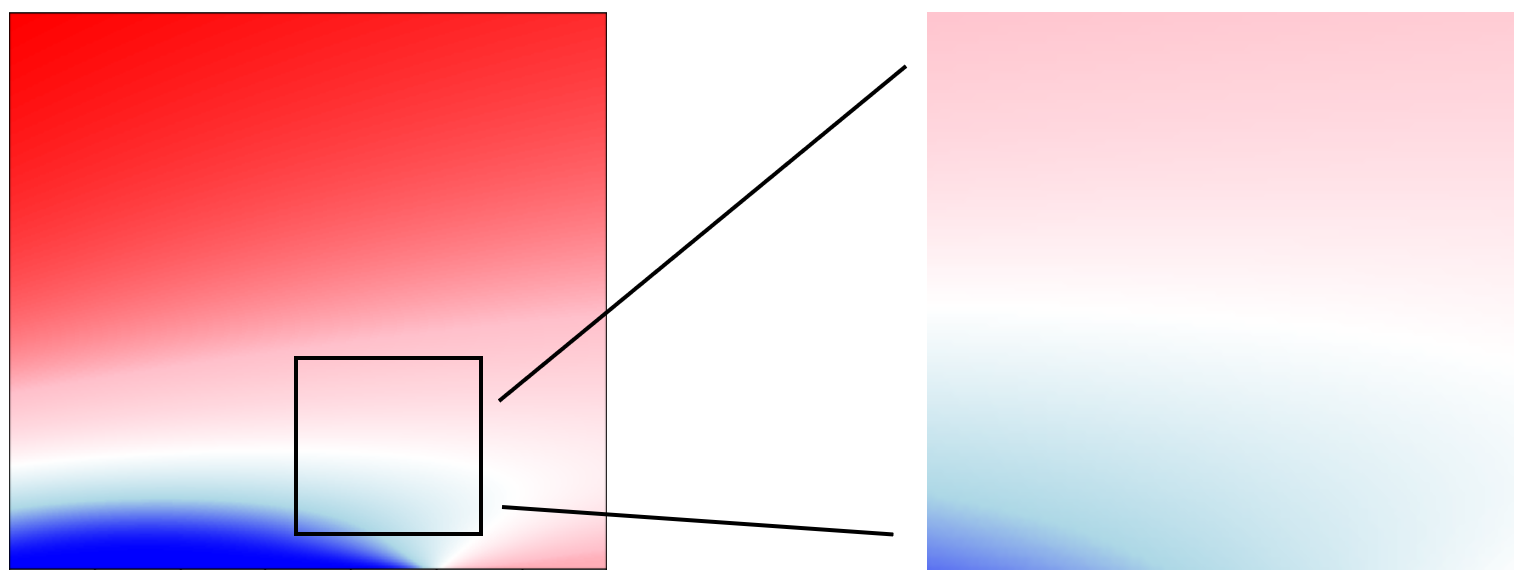}}
    \subfigure[]{ \includegraphics[width=0.8\linewidth]{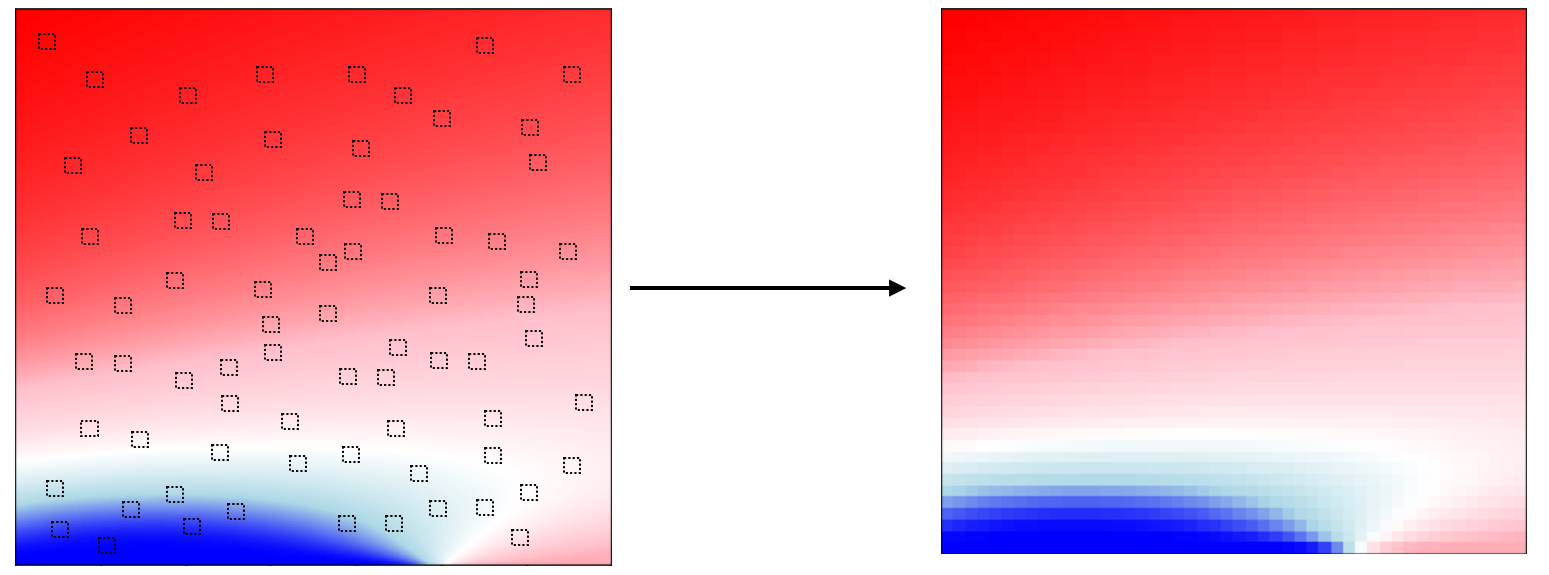}}
    \caption{(a) Cropping a contiguous region of the desired size from a chosen location, with all other information omitted. (b) Randomly sampling the desired number of data points across the entire diagram, thereby preserving most of the overall information.}
    \label{crop}
\end{figure*}

The two cropping strategies employed in this work are also illustrated in Figure \ref{crop}. The first method extracts a contiguous block from the original diagram, thereby including only a portion of the full information in each instance. However, when training is performed in batches over many epochs, this approach effectively covers the entire diagram range, ensuring no loss of representational efficiency. The second method uses random sampling of coordinates, which can capture more diverse information in a single instance but may miss fine structural details. In this study, we employ both cropping strategies in separate stages to strike a balance between coverage and detail, as well as to demonstrate the flexibility of the NN-IQS model.

\section{Detailed derivation of the chiral condensation operator}
The chiral condensation operators in continuous space are first transformed to staggered fermion operators by second quantization. The staggered fermion operators are then further transformed to spin operators by Jordan-Wigner Transformation. Given Dirac fermion operator $\Psi(x) = (\Psi_u(x),\Psi_d(x))^{T}$ and staggered fermion operators $\chi_n,\chi^{\dag}_n$, a detailed derivation of chiral condensation operator $\Bar{\Psi}(x)\Psi(x)$ is provided in~\cite{betavqe,adiabatic}
\begin{equation}
  \frac{\chi_n}{\sqrt{a}} \longleftrightarrow
      \begin{cases}
         \Psi_u(x) & n \text{ : even}\\
         \Psi_d(x) & n \text{ : odd}
      \end{cases}   
\end{equation}
\begin{align}
  \Bar{\Psi}(x)\Psi(x) &= \Psi^{\dag}_u(x)\Psi_u(x) \, - \, \Psi^{\dag}_d(x)\Psi_d(x)\\
                       &= \frac{1}{a}[\chi^{\dag}_{2x}\chi_{2x} \, - \, \chi^{\dag}_{2x-1}\chi_{2x-1}]
\end{align}
Provided translational symmetry~(which may be an approximation in the Schwinger Model we employ given finite lattice dimension),
\begin{align}
  \sum_{x=1}^{N/2}  \Bar{\Psi}(x)\Psi(x) &= \frac{1}{a}\sum_{n=1}^{N}(-1)^{n}\chi^{\dag}_{n}\chi_{n}\\
                                         &= \frac{1}{2a}\sum_{n=1}^{N}(-1)^{n}Z_n
\end{align}
Which further gives expectation value of chiral condensation as
\begin{align}
  \left\langle\Bar{\Psi}(x)\Psi(x) \right\rangle &= \left\langle \psi \middle| \Bar{\Psi}(x)\Psi(x) \middle| \psi \right\rangle\\
                                                 &= \frac{1}{2Na}\left\langle \psi \middle| \sum_{n=1}^{N}(-1)^{n}Z_n \middle| \psi \right\rangle
\end{align}

\section{Feasibility of physical implementation}
\label{app:feasibility_of_physical_implementation}
 Our workflow entails preparing finite-temperature states of the Schwinger model and measuring local observables, with particular emphasis on the chiral condensate $\langle\bar{\Psi}\Psi\rangle=(2Na)^{-1}\sum_{n=1}^{N}(-1)^n\langle Z_n\rangle$. Several strategies exist for preparing Gibbs state on near-term quantum hardware. (i) Variational Thermofield-Double (TFD) protocols purify $e^{-\beta H}/\mathrm{Tr}[e^{-\beta H}]$ on a system--ancilla register and have been analyzed and demonstrated in small-scale experiments~\cite{tfd_prl_2019,tfd_exp_pnas_2020,tfd_npjq_2021}. (ii) Quantum Imaginary-Time Evolution (QITE) approximates $e^{-\beta H}$ with a sequence of shallow, data-driven local unitaries, validated both numerically and on hardware~\cite{qite_natphys_2020,qite_prxq_2021,qite_open_prxq_2022, zhang2021continuous}. (iii) $\beta$-VQE-type variational Gibbs ans\"atze, already applied in Schwinger Model studies, offers a comparable circuit depth alternative~\cite{betavqe}. In all cases the chemical potential appears as a uniform $Z$ field, while temperature scans are realized by varying $\beta$ in the preparation routine.

Upon integrating out the gauge fields, the effective spin Hamiltonian [Eq.~\eqref{Hamiltonian}] features long-range $ZZ$, nearest-neighbor $XX{+}YY$, and staggered on-site $Z$ terms. These ingredients are native to linear trapped-ion simulators with programmable all-to-all couplings. M\o lmer--S\o rensen entangling blocks realize $XX/XY$ interactions; engineered coupling matrices $J_{ij}$, obtained via multi-mode driving and pulse shaping, yield effective $ZZ$ (after basis rotations), while alternating on-site fields encode $(-1)^n$ and the $\mu$ shift through AC-Stark tones. A first-order Trotterization employs blocks $e^{-iH_{XY}\Delta t}$, $e^{-iH_{ZZ}\Delta t}$, and local $Z$ phases; within the parameter window of this work, $K\sim10$--$30$ steps are sufficient to reproduce equilibrium observables within our numerical accuracy. Trapped-ion platforms have already realized real-time Schwinger Model dynamics and general programmable long-range spin interactions, supporting feasibility at $N\!=\!8$--$12$ with extensions toward $N\!\approx\!20$~\cite{martinez_nature_2016,monroe_rmp_2021}.

An analogous decomposition applies to superconducting platforms, where digital circuits implement $XX/YY/ZZ$ blocks with nearest-neighbor gates. Recent large-scale experiments have prepared the Schwinger Model vacua on $\mathcal{O}(10^2)$ qubits using scalable variational ans\"atze, providing a $T{=}0$ reference from which TFD/QITE routines can access finite $T$~\cite{prxq_2024_schwinger100}. Neutral-atom arrays offer a complementary route via constrained Ising/XY dynamics and $U(1)$ quantum-link variants; string-breaking phenomena and related gauge-theory effects have been observed, and local de-tunings implement the staggered fields required here~\cite{zohar_rpp_2016,gross_science_2017}.

Measurements are straightforward: the condensate follows from computational-basis readout of $\{Z_n\}$ with the appropriate staggered sign, while classical-shadow post-processing can reduce the shot cost without altering the circuit structure~\cite{classical_shadow}.

\bibliography{reference}

\begin{thebibliography}{60}%
\makeatletter
\providecommand \@ifxundefined [1]{%
 \@ifx{#1\undefined}
}%
\providecommand \@ifnum [1]{%
 \ifnum #1\expandafter \@firstoftwo
 \else \expandafter \@secondoftwo
 \fi
}%
\providecommand \@ifx [1]{%
 \ifx #1\expandafter \@firstoftwo
 \else \expandafter \@secondoftwo
 \fi
}%
\providecommand \natexlab [1]{#1}%
\providecommand \enquote  [1]{``#1''}%
\providecommand \bibnamefont  [1]{#1}%
\providecommand \bibfnamefont [1]{#1}%
\providecommand \citenamefont [1]{#1}%
\providecommand \href@noop [0]{\@secondoftwo}%
\providecommand \href [0]{\begingroup \@sanitize@url \@href}%
\providecommand \@href[1]{\@@startlink{#1}\@@href}%
\providecommand \@@href[1]{\endgroup#1\@@endlink}%
\providecommand \@sanitize@url [0]{\catcode `\\12\catcode `\$12\catcode
  `\&12\catcode `\#12\catcode `\^12\catcode `\_12\catcode `\%12\relax}%
\providecommand \@@startlink[1]{}%
\providecommand \@@endlink[0]{}%
\providecommand \url  [0]{\begingroup\@sanitize@url \@url }%
\providecommand \@url [1]{\endgroup\@href {#1}{\urlprefix }}%
\providecommand \urlprefix  [0]{URL }%
\providecommand \Eprint [0]{\href }%
\providecommand \doibase [0]{https://doi.org/}%
\providecommand \selectlanguage [0]{\@gobble}%
\providecommand \bibinfo  [0]{\@secondoftwo}%
\providecommand \bibfield  [0]{\@secondoftwo}%
\providecommand \translation [1]{[#1]}%
\providecommand \BibitemOpen [0]{}%
\providecommand \bibitemStop [0]{}%
\providecommand \bibitemNoStop [0]{.\EOS\space}%
\providecommand \EOS [0]{\spacefactor3000\relax}%
\providecommand \BibitemShut  [1]{\csname bibitem#1\endcsname}%
\let\auto@bib@innerbib\@empty
\bibitem [{\citenamefont {Petrova}\ \emph {et~al.}(2024)\citenamefont
  {Petrova}, \citenamefont {Tiunov}, \citenamefont {Bañuls},\ and\
  \citenamefont {Fedorov}}]{schwinger_1}%
  \BibitemOpen
  \bibfield  {author} {\bibinfo {author} {\bibfnamefont {E.~V.}\ \bibnamefont
  {Petrova}}, \bibinfo {author} {\bibfnamefont {E.~S.}\ \bibnamefont {Tiunov}},
  \bibinfo {author} {\bibfnamefont {M.~C.}\ \bibnamefont {Bañuls}},\ and\
  \bibinfo {author} {\bibfnamefont {A.~K.}\ \bibnamefont {Fedorov}},\
  }\href@noop {} {\bibfield  {journal} {\bibinfo  {journal} {Physical review
  letters}\ }\textbf {\bibinfo {volume} {132}},\ \bibinfo {pages} {050401}
  (\bibinfo {year} {2024})}\BibitemShut {NoStop}%
\bibitem [{\citenamefont {Bañuls}\ \emph {et~al.}(2017)\citenamefont
  {Bañuls}, \citenamefont {Cichy}, \citenamefont {Cirac}, \citenamefont
  {Jansen},\ and\ \citenamefont {Kühn}}]{schwinger_2}%
  \BibitemOpen
  \bibfield  {author} {\bibinfo {author} {\bibfnamefont {M.~C.}\ \bibnamefont
  {Bañuls}}, \bibinfo {author} {\bibfnamefont {K.}~\bibnamefont {Cichy}},
  \bibinfo {author} {\bibfnamefont {J.~I.}\ \bibnamefont {Cirac}}, \bibinfo
  {author} {\bibfnamefont {K.}~\bibnamefont {Jansen}},\ and\ \bibinfo {author}
  {\bibfnamefont {S.}~\bibnamefont {Kühn}},\ }\href@noop {} {\bibfield
  {journal} {\bibinfo  {journal} {Physical review letters}\ }\textbf {\bibinfo
  {volume} {118}},\ \bibinfo {pages} {071601} (\bibinfo {year}
  {2017})}\BibitemShut {NoStop}%
\bibitem [{\citenamefont {Belyansky}\ \emph {et~al.}(2024)\citenamefont
  {Belyansky}, \citenamefont {Whitsitt}, \citenamefont {Mueller}, \citenamefont
  {Fahimniya}, \citenamefont {Bennewitz}, \citenamefont {Davoudi},\ and\
  \citenamefont {Gorshkov}}]{schwinger_3}%
  \BibitemOpen
  \bibfield  {author} {\bibinfo {author} {\bibfnamefont {R.}~\bibnamefont
  {Belyansky}}, \bibinfo {author} {\bibfnamefont {S.}~\bibnamefont {Whitsitt}},
  \bibinfo {author} {\bibfnamefont {N.}~\bibnamefont {Mueller}}, \bibinfo
  {author} {\bibfnamefont {A.}~\bibnamefont {Fahimniya}}, \bibinfo {author}
  {\bibfnamefont {E.~R.}\ \bibnamefont {Bennewitz}}, \bibinfo {author}
  {\bibfnamefont {Z.}~\bibnamefont {Davoudi}},\ and\ \bibinfo {author}
  {\bibfnamefont {A.~V.}\ \bibnamefont {Gorshkov}},\ }\href@noop {} {\bibfield
  {journal} {\bibinfo  {journal} {Physical review letters}\ }\textbf {\bibinfo
  {volume} {132}},\ \bibinfo {pages} {091903} (\bibinfo {year}
  {2024})}\BibitemShut {NoStop}%
\bibitem [{\citenamefont {Angelides}\ \emph {et~al.}(2025)\citenamefont
  {Angelides}, \citenamefont {Naredi}, \citenamefont {Crippa}, \citenamefont
  {Jansen}, \citenamefont {Kühn}, \citenamefont {Tavernelli},\ and\
  \citenamefont {Wang}}]{schwinger_4}%
  \BibitemOpen
  \bibfield  {author} {\bibinfo {author} {\bibfnamefont {T.}~\bibnamefont
  {Angelides}}, \bibinfo {author} {\bibfnamefont {P.}~\bibnamefont {Naredi}},
  \bibinfo {author} {\bibfnamefont {A.}~\bibnamefont {Crippa}}, \bibinfo
  {author} {\bibfnamefont {K.}~\bibnamefont {Jansen}}, \bibinfo {author}
  {\bibfnamefont {S.}~\bibnamefont {Kühn}}, \bibinfo {author} {\bibfnamefont
  {I.}~\bibnamefont {Tavernelli}},\ and\ \bibinfo {author} {\bibfnamefont
  {D.~S.}\ \bibnamefont {Wang}},\ }\href@noop {} {\bibfield  {journal}
  {\bibinfo  {journal} {npj quantum information}\ }\textbf {\bibinfo {volume}
  {11}},\ \bibinfo {pages} {6} (\bibinfo {year} {2025})}\BibitemShut {NoStop}%
\bibitem [{\citenamefont {Zache}\ \emph {et~al.}(2019)\citenamefont {Zache},
  \citenamefont {Mueller}, \citenamefont {Schneider}, \citenamefont
  {Jendrzejewski}, \citenamefont {Berges},\ and\ \citenamefont
  {Hauke}}]{schwinger_5}%
  \BibitemOpen
  \bibfield  {author} {\bibinfo {author} {\bibfnamefont {T.}~\bibnamefont
  {Zache}}, \bibinfo {author} {\bibfnamefont {N.}~\bibnamefont {Mueller}},
  \bibinfo {author} {\bibfnamefont {J.}~\bibnamefont {Schneider}}, \bibinfo
  {author} {\bibfnamefont {F.}~\bibnamefont {Jendrzejewski}}, \bibinfo {author}
  {\bibfnamefont {J.}~\bibnamefont {Berges}},\ and\ \bibinfo {author}
  {\bibfnamefont {P.}~\bibnamefont {Hauke}},\ }\href@noop {} {\bibfield
  {journal} {\bibinfo  {journal} {Physical review letters}\ }\textbf {\bibinfo
  {volume} {122}},\ \bibinfo {pages} {050403} (\bibinfo {year}
  {2019})}\BibitemShut {NoStop}%
\bibitem [{\citenamefont {Chanda}\ \emph {et~al.}(2020)\citenamefont {Chanda},
  \citenamefont {Zakrzewski}, \citenamefont {Lewenstein},\ and\ \citenamefont
  {Tagliacozzo}}]{schwinger_6}%
  \BibitemOpen
  \bibfield  {author} {\bibinfo {author} {\bibfnamefont {T.}~\bibnamefont
  {Chanda}}, \bibinfo {author} {\bibfnamefont {J.}~\bibnamefont {Zakrzewski}},
  \bibinfo {author} {\bibfnamefont {M.}~\bibnamefont {Lewenstein}},\ and\
  \bibinfo {author} {\bibfnamefont {L.}~\bibnamefont {Tagliacozzo}},\
  }\href@noop {} {\bibfield  {journal} {\bibinfo  {journal} {Physical review
  letters}\ }\textbf {\bibinfo {volume} {124}},\ \bibinfo {pages} {180602}
  (\bibinfo {year} {2020})}\BibitemShut {NoStop}%
\bibitem [{\citenamefont {Florio}\ \emph {et~al.}(2023)\citenamefont {Florio},
  \citenamefont {Frenklakh}, \citenamefont {Ikeda}, \citenamefont {Kharzeev},
  \citenamefont {Korepin}, \citenamefont {Shi},\ and\ \citenamefont
  {Yu}}]{schwinger_7}%
  \BibitemOpen
  \bibfield  {author} {\bibinfo {author} {\bibfnamefont {A.}~\bibnamefont
  {Florio}}, \bibinfo {author} {\bibfnamefont {D.}~\bibnamefont {Frenklakh}},
  \bibinfo {author} {\bibfnamefont {K.}~\bibnamefont {Ikeda}}, \bibinfo
  {author} {\bibfnamefont {D.}~\bibnamefont {Kharzeev}}, \bibinfo {author}
  {\bibfnamefont {V.}~\bibnamefont {Korepin}}, \bibinfo {author} {\bibfnamefont
  {S.}~\bibnamefont {Shi}},\ and\ \bibinfo {author} {\bibfnamefont
  {K.}~\bibnamefont {Yu}},\ }\href@noop {} {\bibfield  {journal} {\bibinfo
  {journal} {Physical review letters}\ }\textbf {\bibinfo {volume} {131}},\
  \bibinfo {pages} {021902} (\bibinfo {year} {2023})}\BibitemShut {NoStop}%
\bibitem [{\citenamefont {Dempsey}\ \emph
  {et~al.}(2024{\natexlab{a}})\citenamefont {Dempsey}, \citenamefont
  {Klebanov}, \citenamefont {Pufu}, \citenamefont {Søgaard},\ and\
  \citenamefont {Zan}}]{schwinger_8}%
  \BibitemOpen
  \bibfield  {author} {\bibinfo {author} {\bibfnamefont {R.}~\bibnamefont
  {Dempsey}}, \bibinfo {author} {\bibfnamefont {I.~R.}\ \bibnamefont
  {Klebanov}}, \bibinfo {author} {\bibfnamefont {S.~S.}\ \bibnamefont {Pufu}},
  \bibinfo {author} {\bibfnamefont {B.~T.}\ \bibnamefont {Søgaard}},\ and\
  \bibinfo {author} {\bibfnamefont {B.}~\bibnamefont {Zan}},\ }\href@noop {}
  {\bibfield  {journal} {\bibinfo  {journal} {Physical review letters}\
  }\textbf {\bibinfo {volume} {132}},\ \bibinfo {pages} {031603} (\bibinfo
  {year} {2024}{\natexlab{a}})}\BibitemShut {NoStop}%
\bibitem [{\citenamefont {Philipsen}(2021)}]{symmetry1302079}%
  \BibitemOpen
  \bibfield  {author} {\bibinfo {author} {\bibfnamefont {O.}~\bibnamefont
  {Philipsen}},\ }\href@noop {} {\bibfield  {journal} {\bibinfo  {journal}
  {Symmetry (Basel)}\ }\textbf {\bibinfo {volume} {13}},\ \bibinfo {pages}
  {2079} (\bibinfo {year} {2021})}\BibitemShut {NoStop}%
\bibitem [{\citenamefont {Stephanov}(2005)}]{Stephanov2012}%
  \BibitemOpen
  \bibfield  {author} {\bibinfo {author} {\bibfnamefont {M.~A.}\ \bibnamefont
  {Stephanov}},\ }\href@noop {} {\bibfield  {journal} {\bibinfo  {journal}
  {International journal of modern physics. A, Particles and fields,
  gravitation, cosmology}\ }\textbf {\bibinfo {volume} {20}},\ \bibinfo {pages}
  {4387} (\bibinfo {year} {2005})}\BibitemShut {NoStop}%
\bibitem [{\citenamefont {Guenther}(2022)}]{LATTICE2021}%
  \BibitemOpen
  \bibfield  {author} {\bibinfo {author} {\bibfnamefont {J.~N.}\ \bibnamefont
  {Guenther}},\ }\href@noop {} {\bibfield  {journal} {\bibinfo  {journal}
  {arXiv.org}\ } (\bibinfo {year} {2022})}\BibitemShut {NoStop}%
\bibitem [{\citenamefont {Fu}\ \emph {et~al.}(2020)\citenamefont {Fu},
  \citenamefont {Pawlowski},\ and\ \citenamefont
  {Rennecke}}]{PHysRevD101054032}%
  \BibitemOpen
  \bibfield  {author} {\bibinfo {author} {\bibfnamefont {W.-j.}\ \bibnamefont
  {Fu}}, \bibinfo {author} {\bibfnamefont {J.~M.}\ \bibnamefont {Pawlowski}},\
  and\ \bibinfo {author} {\bibfnamefont {F.}~\bibnamefont {Rennecke}},\
  }\href@noop {} {\bibfield  {journal} {\bibinfo  {journal} {Physical review.
  D}\ }\textbf {\bibinfo {volume} {101}},\ \bibinfo {pages} {1} (\bibinfo
  {year} {2020})}\BibitemShut {NoStop}%
\bibitem [{\citenamefont {Ratti}(2018)}]{heavyion}%
  \BibitemOpen
  \bibfield  {author} {\bibinfo {author} {\bibfnamefont {C.}~\bibnamefont
  {Ratti}},\ }\href@noop {} {\bibfield  {journal} {\bibinfo  {journal} {Reports
  on progress in physics}\ }\textbf {\bibinfo {volume} {81}},\ \bibinfo {pages}
  {84301} (\bibinfo {year} {2018})}\BibitemShut {NoStop}%
\bibitem [{\citenamefont {Ding}(2021)}]{200211957v1}%
  \BibitemOpen
  \bibfield  {author} {\bibinfo {author} {\bibfnamefont {H.-T.}\ \bibnamefont
  {Ding}},\ }\href@noop {} {\bibfield  {journal} {\bibinfo  {journal} {Nuclear
  physics. A}\ }\textbf {\bibinfo {volume} {1005}},\ \bibinfo {pages} {121940}
  (\bibinfo {year} {2021})}\BibitemShut {NoStop}%
\bibitem [{\citenamefont {Ohata}(2024)}]{hirokiohata}%
  \BibitemOpen
  \bibfield  {author} {\bibinfo {author} {\bibfnamefont {H.}~\bibnamefont
  {Ohata}},\ }\href@noop {} {\bibfield  {journal} {\bibinfo  {journal}
  {Progress of theoretical and experimental physics}\ }\textbf {\bibinfo
  {volume} {2024}} (\bibinfo {year} {2024})}\BibitemShut {NoStop}%
\bibitem [{\citenamefont {Li}\ \emph {et~al.}(2024)\citenamefont {Li},
  \citenamefont {Li}, \citenamefont {Zhuang},\ and\ \citenamefont
  {Yung}}]{xiaowei}%
  \BibitemOpen
  \bibfield  {author} {\bibinfo {author} {\bibfnamefont {X.-W.}\ \bibnamefont
  {Li}}, \bibinfo {author} {\bibfnamefont {F.}~\bibnamefont {Li}}, \bibinfo
  {author} {\bibfnamefont {J.}~\bibnamefont {Zhuang}},\ and\ \bibinfo {author}
  {\bibfnamefont {M.-H.}\ \bibnamefont {Yung}},\ }\href@noop {} {\bibfield
  {journal} {\bibinfo  {journal} {arXiv.org}\ } (\bibinfo {year}
  {2024})}\BibitemShut {NoStop}%
\bibitem [{\citenamefont {Dempsey}\ \emph
  {et~al.}(2024{\natexlab{b}})\citenamefont {Dempsey}, \citenamefont
  {Klebanov}, \citenamefont {Pufu}, \citenamefont {Søgaard},\ and\
  \citenamefont {Zan}}]{rossdempsey}%
  \BibitemOpen
  \bibfield  {author} {\bibinfo {author} {\bibfnamefont {R.}~\bibnamefont
  {Dempsey}}, \bibinfo {author} {\bibfnamefont {I.~R.}\ \bibnamefont
  {Klebanov}}, \bibinfo {author} {\bibfnamefont {S.~S.}\ \bibnamefont {Pufu}},
  \bibinfo {author} {\bibfnamefont {B.~T.}\ \bibnamefont {Søgaard}},\ and\
  \bibinfo {author} {\bibfnamefont {B.}~\bibnamefont {Zan}},\ }\href@noop {}
  {\bibfield  {journal} {\bibinfo  {journal} {Physical review letters}\
  }\textbf {\bibinfo {volume} {132}},\ \bibinfo {pages} {031603} (\bibinfo
  {year} {2024}{\natexlab{b}})}\BibitemShut {NoStop}%
\bibitem [{\citenamefont {Shimizu}\ and\ \citenamefont
  {Kuramashi}(2014)}]{yuyashimizu}%
  \BibitemOpen
  \bibfield  {author} {\bibinfo {author} {\bibfnamefont {Y.}~\bibnamefont
  {Shimizu}}\ and\ \bibinfo {author} {\bibfnamefont {Y.}~\bibnamefont
  {Kuramashi}},\ }\href@noop {} {\bibfield  {journal} {\bibinfo  {journal}
  {Physical review. D, Particles, fields, gravitation, and cosmology}\ }\textbf
  {\bibinfo {volume} {90}} (\bibinfo {year} {2014})}\BibitemShut {NoStop}%
\bibitem [{\citenamefont {Butt}\ \emph {et~al.}(2020)\citenamefont {Butt},
  \citenamefont {Catterall}, \citenamefont {Meurice}, \citenamefont {Sakai},\
  and\ \citenamefont {Unmuth-Yockey}}]{nbutt}%
  \BibitemOpen
  \bibfield  {author} {\bibinfo {author} {\bibfnamefont {N.}~\bibnamefont
  {Butt}}, \bibinfo {author} {\bibfnamefont {S.}~\bibnamefont {Catterall}},
  \bibinfo {author} {\bibfnamefont {Y.}~\bibnamefont {Meurice}}, \bibinfo
  {author} {\bibfnamefont {R.}~\bibnamefont {Sakai}},\ and\ \bibinfo {author}
  {\bibfnamefont {J.}~\bibnamefont {Unmuth-Yockey}},\ }\href@noop {} {\bibfield
   {journal} {\bibinfo  {journal} {Physical review. D}\ }\textbf {\bibinfo
  {volume} {101}} (\bibinfo {year} {2020})}\BibitemShut {NoStop}%
\bibitem [{\citenamefont {Banuls}\ \emph {et~al.}(2017)\citenamefont {Banuls},
  \citenamefont {Cichy}, \citenamefont {Cirac}, \citenamefont {Jansen},\ and\
  \citenamefont {Kuehn}}]{mbanuls}%
  \BibitemOpen
  \bibfield  {author} {\bibinfo {author} {\bibfnamefont {M.~C.}\ \bibnamefont
  {Banuls}}, \bibinfo {author} {\bibfnamefont {K.}~\bibnamefont {Cichy}},
  \bibinfo {author} {\bibfnamefont {J.~I.}\ \bibnamefont {Cirac}}, \bibinfo
  {author} {\bibfnamefont {K.}~\bibnamefont {Jansen}},\ and\ \bibinfo {author}
  {\bibfnamefont {S.}~\bibnamefont {Kuehn}},\ }\href@noop {} {\bibfield
  {journal} {\bibinfo  {journal} {Physical review letters}\ }\textbf {\bibinfo
  {volume} {118}},\ \bibinfo {pages} {071601} (\bibinfo {year}
  {2017})}\BibitemShut {NoStop}%
\bibitem [{\citenamefont {Rico}\ \emph {et~al.}(2014)\citenamefont {Rico},
  \citenamefont {Pichler}, \citenamefont {Dalmonte}, \citenamefont {Zoller},\
  and\ \citenamefont {Montangero}}]{erico}%
  \BibitemOpen
  \bibfield  {author} {\bibinfo {author} {\bibfnamefont {E.}~\bibnamefont
  {Rico}}, \bibinfo {author} {\bibfnamefont {T.}~\bibnamefont {Pichler}},
  \bibinfo {author} {\bibfnamefont {M.}~\bibnamefont {Dalmonte}}, \bibinfo
  {author} {\bibfnamefont {P.}~\bibnamefont {Zoller}},\ and\ \bibinfo {author}
  {\bibfnamefont {S.}~\bibnamefont {Montangero}},\ }\href@noop {} {\bibfield
  {journal} {\bibinfo  {journal} {Physical review letters}\ }\textbf {\bibinfo
  {volume} {112}} (\bibinfo {year} {2014})}\BibitemShut {NoStop}%
\bibitem [{\citenamefont {Fujii}\ \emph {et~al.}(2024)\citenamefont {Fujii},
  \citenamefont {Fujikura}, \citenamefont {Kikukawa}, \citenamefont {Okuda},\
  and\ \citenamefont {Pedersen}}]{hirotsugufujii}%
  \BibitemOpen
  \bibfield  {author} {\bibinfo {author} {\bibfnamefont {H.}~\bibnamefont
  {Fujii}}, \bibinfo {author} {\bibfnamefont {K.}~\bibnamefont {Fujikura}},
  \bibinfo {author} {\bibfnamefont {Y.}~\bibnamefont {Kikukawa}}, \bibinfo
  {author} {\bibfnamefont {T.}~\bibnamefont {Okuda}},\ and\ \bibinfo {author}
  {\bibfnamefont {J.~W.}\ \bibnamefont {Pedersen}},\ }\href@noop {} {\bibfield
  {journal} {\bibinfo  {journal} {arXiv.org}\ } (\bibinfo {year}
  {2024})}\BibitemShut {NoStop}%
\bibitem [{\citenamefont {Funcke}\ \emph {et~al.}(2020)\citenamefont {Funcke},
  \citenamefont {Jansen},\ and\ \citenamefont {Kühn}}]{lfuncke}%
  \BibitemOpen
  \bibfield  {author} {\bibinfo {author} {\bibfnamefont {L.}~\bibnamefont
  {Funcke}}, \bibinfo {author} {\bibfnamefont {K.}~\bibnamefont {Jansen}},\
  and\ \bibinfo {author} {\bibfnamefont {S.}~\bibnamefont {Kühn}},\
  }\href@noop {} {\bibfield  {journal} {\bibinfo  {journal} {Physical review.
  D}\ }\textbf {\bibinfo {volume} {101}},\ \bibinfo {pages} {1} (\bibinfo
  {year} {2020})}\BibitemShut {NoStop}%
\bibitem [{\citenamefont {Byrnes}\ \emph
  {et~al.}(2002{\natexlab{a}})\citenamefont {Byrnes}, \citenamefont
  {Sriganesh}, \citenamefont {Bursill},\ and\ \citenamefont {Hamer}}]{tbyrnes}%
  \BibitemOpen
  \bibfield  {author} {\bibinfo {author} {\bibfnamefont {T.}~\bibnamefont
  {Byrnes}}, \bibinfo {author} {\bibfnamefont {P.}~\bibnamefont {Sriganesh}},
  \bibinfo {author} {\bibfnamefont {R.}~\bibnamefont {Bursill}},\ and\ \bibinfo
  {author} {\bibfnamefont {C.}~\bibnamefont {Hamer}},\ }\href@noop {}
  {\bibfield  {journal} {\bibinfo  {journal} {Nuclear physics. Section B,
  Proceedings supplement}\ }\textbf {\bibinfo {volume} {109}},\ \bibinfo
  {pages} {202} (\bibinfo {year} {2002}{\natexlab{a}})}\BibitemShut {NoStop}%
\bibitem [{\citenamefont {Jónsson}\ \emph {et~al.}(2018)\citenamefont
  {Jónsson}, \citenamefont {Bauer},\ and\ \citenamefont {Carleo}}]{180805232}%
  \BibitemOpen
  \bibfield  {author} {\bibinfo {author} {\bibfnamefont {B.}~\bibnamefont
  {Jónsson}}, \bibinfo {author} {\bibfnamefont {B.}~\bibnamefont {Bauer}},\
  and\ \bibinfo {author} {\bibfnamefont {G.}~\bibnamefont {Carleo}},\
  }\href@noop {} {\bibfield  {journal} {\bibinfo  {journal} {arXiv.org}\ }
  (\bibinfo {year} {2018})}\BibitemShut {NoStop}%
\bibitem [{\citenamefont {Díaz-Pier}\ \emph {et~al.}(2011)\citenamefont
  {Díaz-Pier}, \citenamefont {Venegas-Andraca},\ and\ \citenamefont
  {Gómez-Muñoz}}]{11031399}%
  \BibitemOpen
  \bibfield  {author} {\bibinfo {author} {\bibfnamefont {S.}~\bibnamefont
  {Díaz-Pier}}, \bibinfo {author} {\bibfnamefont {S.~E.}\ \bibnamefont
  {Venegas-Andraca}},\ and\ \bibinfo {author} {\bibfnamefont {J.~L.}\
  \bibnamefont {Gómez-Muñoz}},\ }\href@noop {} {\bibfield  {journal}
  {\bibinfo  {journal} {arXiv.org}\ } (\bibinfo {year} {2011})}\BibitemShut
  {NoStop}%
\bibitem [{\citenamefont {Guerreschi}\ \emph {et~al.}(2020)\citenamefont
  {Guerreschi}, \citenamefont {Hogaboam}, \citenamefont {Baruffa},\ and\
  \citenamefont {Sawaya}}]{guerreschi}%
  \BibitemOpen
  \bibfield  {author} {\bibinfo {author} {\bibfnamefont {G.~G.}\ \bibnamefont
  {Guerreschi}}, \bibinfo {author} {\bibfnamefont {J.}~\bibnamefont
  {Hogaboam}}, \bibinfo {author} {\bibfnamefont {F.}~\bibnamefont {Baruffa}},\
  and\ \bibinfo {author} {\bibfnamefont {N.~P.~D.}\ \bibnamefont {Sawaya}},\
  }\href@noop {} {\bibfield  {journal} {\bibinfo  {journal} {Quantum science
  and technology}\ }\textbf {\bibinfo {volume} {5}},\ \bibinfo {pages} {34007}
  (\bibinfo {year} {2020})}\BibitemShut {NoStop}%
\bibitem [{\citenamefont {Vallero}\ \emph {et~al.}(2026)\citenamefont
  {Vallero}, \citenamefont {Rech},\ and\ \citenamefont {Vella}}]{1s20s0167}%
  \BibitemOpen
  \bibfield  {author} {\bibinfo {author} {\bibfnamefont {M.}~\bibnamefont
  {Vallero}}, \bibinfo {author} {\bibfnamefont {P.}~\bibnamefont {Rech}},\ and\
  \bibinfo {author} {\bibfnamefont {F.}~\bibnamefont {Vella}},\ }\href@noop {}
  {\bibfield  {journal} {\bibinfo  {journal} {Future generation computer
  systems}\ }\textbf {\bibinfo {volume} {174}} (\bibinfo {year}
  {2026})}\BibitemShut {NoStop}%
\bibitem [{\citenamefont {De~Raedt}\ \emph {et~al.}(2019)\citenamefont
  {De~Raedt}, \citenamefont {Jin}, \citenamefont {Willsch}, \citenamefont
  {Willsch}, \citenamefont {Yoshioka}, \citenamefont {Ito}, \citenamefont
  {Yuan},\ and\ \citenamefont {Michielsen}}]{1s20s001046}%
  \BibitemOpen
  \bibfield  {author} {\bibinfo {author} {\bibfnamefont {H.}~\bibnamefont
  {De~Raedt}}, \bibinfo {author} {\bibfnamefont {F.}~\bibnamefont {Jin}},
  \bibinfo {author} {\bibfnamefont {D.}~\bibnamefont {Willsch}}, \bibinfo
  {author} {\bibfnamefont {M.}~\bibnamefont {Willsch}}, \bibinfo {author}
  {\bibfnamefont {N.}~\bibnamefont {Yoshioka}}, \bibinfo {author}
  {\bibfnamefont {N.}~\bibnamefont {Ito}}, \bibinfo {author} {\bibfnamefont
  {S.}~\bibnamefont {Yuan}},\ and\ \bibinfo {author} {\bibfnamefont
  {K.}~\bibnamefont {Michielsen}},\ }\href@noop {} {\bibfield  {journal}
  {\bibinfo  {journal} {Computer physics communications}\ }\textbf {\bibinfo
  {volume} {237}},\ \bibinfo {pages} {47} (\bibinfo {year} {2019})}\BibitemShut
  {NoStop}%
\bibitem [{\citenamefont {Bravyi}\ \emph {et~al.}(2016)\citenamefont {Bravyi},
  \citenamefont {Smith},\ and\ \citenamefont {Smolin}}]{physrevx1004}%
  \BibitemOpen
  \bibfield  {author} {\bibinfo {author} {\bibfnamefont {S.}~\bibnamefont
  {Bravyi}}, \bibinfo {author} {\bibfnamefont {G.}~\bibnamefont {Smith}},\ and\
  \bibinfo {author} {\bibfnamefont {J.~A.}\ \bibnamefont {Smolin}},\
  }\href@noop {} {\bibfield  {journal} {\bibinfo  {journal} {Physical review.
  X}\ }\textbf {\bibinfo {volume} {6}},\ \bibinfo {pages} {021043} (\bibinfo
  {year} {2016})}\BibitemShut {NoStop}%
\bibitem [{\citenamefont {Fösel}\ \emph {et~al.}(2018)\citenamefont {Fösel},
  \citenamefont {Tighineanu}, \citenamefont {Weiss},\ and\ \citenamefont
  {Marquardt}}]{physrevx8031084}%
  \BibitemOpen
  \bibfield  {author} {\bibinfo {author} {\bibfnamefont {T.}~\bibnamefont
  {Fösel}}, \bibinfo {author} {\bibfnamefont {P.}~\bibnamefont {Tighineanu}},
  \bibinfo {author} {\bibfnamefont {T.}~\bibnamefont {Weiss}},\ and\ \bibinfo
  {author} {\bibfnamefont {F.}~\bibnamefont {Marquardt}},\ }\href@noop {}
  {\bibfield  {journal} {\bibinfo  {journal} {Physical review. X}\ }\textbf
  {\bibinfo {volume} {8}},\ \bibinfo {pages} {031084} (\bibinfo {year}
  {2018})}\BibitemShut {NoStop}%
\bibitem [{\citenamefont {Marquardt}(2021)}]{scipostphyslectnotes29}%
  \BibitemOpen
  \bibfield  {author} {\bibinfo {author} {\bibfnamefont {F.}~\bibnamefont
  {Marquardt}},\ }\href@noop {} {\bibfield  {journal} {\bibinfo  {journal}
  {SciPost physics lecture notes}\ ,\ \bibinfo {pages} {29}} (\bibinfo {year}
  {2021})}\BibitemShut {NoStop}%
\bibitem [{\citenamefont {Lamata}(2021)}]{photonics0800033}%
  \BibitemOpen
  \bibfield  {author} {\bibinfo {author} {\bibfnamefont {L.}~\bibnamefont
  {Lamata}},\ }\href@noop {} {\bibfield  {journal} {\bibinfo  {journal}
  {Photonics}\ }\textbf {\bibinfo {volume} {8}},\ \bibinfo {pages} {33}
  (\bibinfo {year} {2021})}\BibitemShut {NoStop}%
\bibitem [{\citenamefont {Gebhart}\ \emph {et~al.}(2023)\citenamefont
  {Gebhart}, \citenamefont {Santagati}, \citenamefont {Gentile}, \citenamefont
  {Gauger}, \citenamefont {Craig}, \citenamefont {Ares}, \citenamefont
  {Banchi}, \citenamefont {Marquardt}, \citenamefont {Pezzè},\ and\
  \citenamefont {Bonato}}]{220700298v3}%
  \BibitemOpen
  \bibfield  {author} {\bibinfo {author} {\bibfnamefont {V.}~\bibnamefont
  {Gebhart}}, \bibinfo {author} {\bibfnamefont {R.}~\bibnamefont {Santagati}},
  \bibinfo {author} {\bibfnamefont {A.~A.}\ \bibnamefont {Gentile}}, \bibinfo
  {author} {\bibfnamefont {E.~M.}\ \bibnamefont {Gauger}}, \bibinfo {author}
  {\bibfnamefont {D.}~\bibnamefont {Craig}}, \bibinfo {author} {\bibfnamefont
  {N.}~\bibnamefont {Ares}}, \bibinfo {author} {\bibfnamefont {L.}~\bibnamefont
  {Banchi}}, \bibinfo {author} {\bibfnamefont {F.}~\bibnamefont {Marquardt}},
  \bibinfo {author} {\bibfnamefont {L.}~\bibnamefont {Pezzè}},\ and\ \bibinfo
  {author} {\bibfnamefont {C.}~\bibnamefont {Bonato}},\ }\href@noop {}
  {\bibfield  {journal} {\bibinfo  {journal} {Nature reviews physics}\ }\textbf
  {\bibinfo {volume} {5}},\ \bibinfo {pages} {141} (\bibinfo {year}
  {2023})}\BibitemShut {NoStop}%
\bibitem [{\citenamefont {Krenn}\ \emph {et~al.}(2022)\citenamefont {Krenn},
  \citenamefont {Landgraf}, \citenamefont {Foesel},\ and\ \citenamefont
  {Marquardt}}]{physreva107010101}%
  \BibitemOpen
  \bibfield  {author} {\bibinfo {author} {\bibfnamefont {M.}~\bibnamefont
  {Krenn}}, \bibinfo {author} {\bibfnamefont {J.}~\bibnamefont {Landgraf}},
  \bibinfo {author} {\bibfnamefont {T.}~\bibnamefont {Foesel}},\ and\ \bibinfo
  {author} {\bibfnamefont {F.}~\bibnamefont {Marquardt}},\ }\href@noop {}
  {\bibfield  {journal} {\bibinfo  {journal} {arXiv.org}\ } (\bibinfo {year}
  {2022})}\BibitemShut {NoStop}%
\bibitem [{\citenamefont {Zhu}\ \emph {et~al.}(2022)\citenamefont {Zhu},
  \citenamefont {Wu}, \citenamefont {Bai}, \citenamefont {Wang}, \citenamefont
  {Wang},\ and\ \citenamefont {Chiribella}}]{replearn1}%
  \BibitemOpen
  \bibfield  {author} {\bibinfo {author} {\bibfnamefont {Y.}~\bibnamefont
  {Zhu}}, \bibinfo {author} {\bibfnamefont {Y.-D.}\ \bibnamefont {Wu}},
  \bibinfo {author} {\bibfnamefont {G.}~\bibnamefont {Bai}}, \bibinfo {author}
  {\bibfnamefont {D.-S.}\ \bibnamefont {Wang}}, \bibinfo {author}
  {\bibfnamefont {Y.}~\bibnamefont {Wang}},\ and\ \bibinfo {author}
  {\bibfnamefont {G.}~\bibnamefont {Chiribella}},\ }\href@noop {} {\bibfield
  {journal} {\bibinfo  {journal} {Nature communications}\ }\textbf {\bibinfo
  {volume} {13}},\ \bibinfo {pages} {6222} (\bibinfo {year}
  {2022})}\BibitemShut {NoStop}%
\bibitem [{\citenamefont {Gao}\ and\ \citenamefont {Duan}(2017)}]{replearn2}%
  \BibitemOpen
  \bibfield  {author} {\bibinfo {author} {\bibfnamefont {X.}~\bibnamefont
  {Gao}}\ and\ \bibinfo {author} {\bibfnamefont {L.-M.}\ \bibnamefont {Duan}},\
  }\href@noop {} {\bibfield  {journal} {\bibinfo  {journal} {Nature
  communications}\ }\textbf {\bibinfo {volume} {8}},\ \bibinfo {pages} {662}
  (\bibinfo {year} {2017})}\BibitemShut {NoStop}%
\bibitem [{\citenamefont {Carleo}\ and\ \citenamefont
  {Troyer}(2017)}]{replearn3}%
  \BibitemOpen
  \bibfield  {author} {\bibinfo {author} {\bibfnamefont {G.}~\bibnamefont
  {Carleo}}\ and\ \bibinfo {author} {\bibfnamefont {M.}~\bibnamefont
  {Troyer}},\ }\href@noop {} {\bibfield  {journal} {\bibinfo  {journal}
  {Science (American Association for the Advancement of Science)}\ }\textbf
  {\bibinfo {volume} {355}},\ \bibinfo {pages} {602} (\bibinfo {year}
  {2017})}\BibitemShut {NoStop}%
\bibitem [{\citenamefont {Ahmed}\ \emph {et~al.}(2022)\citenamefont {Ahmed},
  \citenamefont {Killoran},\ and\ \citenamefont {Álvarez}}]{221113765v1}%
  \BibitemOpen
  \bibfield  {author} {\bibinfo {author} {\bibfnamefont {S.}~\bibnamefont
  {Ahmed}}, \bibinfo {author} {\bibfnamefont {N.}~\bibnamefont {Killoran}},\
  and\ \bibinfo {author} {\bibfnamefont {J.~F.~C.}\ \bibnamefont {Álvarez}},\
  }\href@noop {} {\bibfield  {journal} {\bibinfo  {journal} {arXiv.org}\ }
  (\bibinfo {year} {2022})}\BibitemShut {NoStop}%
\bibitem [{\citenamefont {Tomiya}(2022)}]{betavqe}%
  \BibitemOpen
  \bibfield  {author} {\bibinfo {author} {\bibfnamefont {A.}~\bibnamefont
  {Tomiya}},\ }\href@noop {} {\bibfield  {journal} {\bibinfo  {journal} {arXiv
  preprint arXiv:2205.08860}\ } (\bibinfo {year} {2022})}\BibitemShut {NoStop}%
\bibitem [{\citenamefont {Chakraborty}\ \emph {et~al.}(2022)\citenamefont
  {Chakraborty}, \citenamefont {Honda}, \citenamefont {Izubuchi}, \citenamefont
  {Kikuchi},\ and\ \citenamefont {Tomiya}}]{adiabatic}%
  \BibitemOpen
  \bibfield  {author} {\bibinfo {author} {\bibfnamefont {B.}~\bibnamefont
  {Chakraborty}}, \bibinfo {author} {\bibfnamefont {M.}~\bibnamefont {Honda}},
  \bibinfo {author} {\bibfnamefont {T.}~\bibnamefont {Izubuchi}}, \bibinfo
  {author} {\bibfnamefont {Y.}~\bibnamefont {Kikuchi}},\ and\ \bibinfo {author}
  {\bibfnamefont {A.}~\bibnamefont {Tomiya}},\ }\href@noop {} {\bibfield
  {journal} {\bibinfo  {journal} {Physical review. D}\ }\textbf {\bibinfo
  {volume} {105}} (\bibinfo {year} {2022})}\BibitemShut {NoStop}%
\bibitem [{\citenamefont {Jentsch}\ \emph {et~al.}(2022)\citenamefont
  {Jentsch}, \citenamefont {Daviet}, \citenamefont {Dupuis},\ and\
  \citenamefont {Floerchinger}}]{schwingertransition}%
  \BibitemOpen
  \bibfield  {author} {\bibinfo {author} {\bibfnamefont {P.}~\bibnamefont
  {Jentsch}}, \bibinfo {author} {\bibfnamefont {R.}~\bibnamefont {Daviet}},
  \bibinfo {author} {\bibfnamefont {N.}~\bibnamefont {Dupuis}},\ and\ \bibinfo
  {author} {\bibfnamefont {S.}~\bibnamefont {Floerchinger}},\ }\href@noop {}
  {\bibfield  {journal} {\bibinfo  {journal} {Physical review. D}\ }\textbf
  {\bibinfo {volume} {105}} (\bibinfo {year} {2022})}\BibitemShut {NoStop}%
\bibitem [{\citenamefont {Wipf}(2021)}]{theory}%
  \BibitemOpen
  \bibfield  {author} {\bibinfo {author} {\bibfnamefont {A.}~\bibnamefont
  {Wipf}},\ }\href@noop {} {{\selectlanguage {eng}\emph {\bibinfo {title}
  {Statistical approach to quantum field theory : an introduction}}}},\
  \bibinfo {edition} {second edition.}\ ed.,\ Lecture notes in physics, volume
  992\ (\bibinfo  {publisher} {Springer},\ \bibinfo {address} {Cham},\ \bibinfo
  {year} {2021})\BibitemShut {NoStop}%
\bibitem [{\citenamefont {Chen}\ \emph {et~al.}(2021)\citenamefont {Chen},
  \citenamefont {Liu},\ and\ \citenamefont {Wang}}]{LIIF}%
  \BibitemOpen
  \bibfield  {author} {\bibinfo {author} {\bibfnamefont {Y.}~\bibnamefont
  {Chen}}, \bibinfo {author} {\bibfnamefont {S.}~\bibnamefont {Liu}},\ and\
  \bibinfo {author} {\bibfnamefont {X.}~\bibnamefont {Wang}},\ }in\ \href@noop
  {} {\emph {\bibinfo {booktitle} {Proceedings of the IEEE/CVF Conference on
  Computer Vision and Pattern Recognition}}}\ (\bibinfo {year} {2021})\ pp.\
  \bibinfo {pages} {8628--8638}\BibitemShut {NoStop}%
\bibitem [{\citenamefont {Byrnes}\ \emph
  {et~al.}(2002{\natexlab{b}})\citenamefont {Byrnes}, \citenamefont
  {Sriganesh}, \citenamefont {Bursill},\ and\ \citenamefont
  {Hamer}}]{transition1}%
  \BibitemOpen
  \bibfield  {author} {\bibinfo {author} {\bibfnamefont {T.~M.~R.}\
  \bibnamefont {Byrnes}}, \bibinfo {author} {\bibfnamefont {P.}~\bibnamefont
  {Sriganesh}}, \bibinfo {author} {\bibfnamefont {R.~J.}\ \bibnamefont
  {Bursill}},\ and\ \bibinfo {author} {\bibfnamefont {C.~J.}\ \bibnamefont
  {Hamer}},\ }\href@noop {} {\bibfield  {journal} {\bibinfo  {journal}
  {Physical review. D, Particles and fields}\ }\textbf {\bibinfo {volume} {66}}
  (\bibinfo {year} {2002}{\natexlab{b}})}\BibitemShut {NoStop}%
\bibitem [{\citenamefont {Bañuls}\ \emph {et~al.}(2018)\citenamefont
  {Bañuls}, \citenamefont {Cichy}, \citenamefont {Cirac}, \citenamefont
  {Jansen},\ and\ \citenamefont {Kühn}}]{transition2}%
  \BibitemOpen
  \bibfield  {author} {\bibinfo {author} {\bibfnamefont {M.~C.}\ \bibnamefont
  {Bañuls}}, \bibinfo {author} {\bibfnamefont {K.}~\bibnamefont {Cichy}},
  \bibinfo {author} {\bibfnamefont {J.~I.}\ \bibnamefont {Cirac}}, \bibinfo
  {author} {\bibfnamefont {K.}~\bibnamefont {Jansen}},\ and\ \bibinfo {author}
  {\bibfnamefont {S.}~\bibnamefont {Kühn}},\ }\href@noop {} {\bibfield
  {journal} {\bibinfo  {journal} {arXiv.org}\ } (\bibinfo {year}
  {2018})}\BibitemShut {NoStop}%
\bibitem [{\citenamefont {Troyer}\ and\ \citenamefont
  {Wiese}(2005)}]{signprob1}%
  \BibitemOpen
  \bibfield  {author} {\bibinfo {author} {\bibfnamefont {M.}~\bibnamefont
  {Troyer}}\ and\ \bibinfo {author} {\bibfnamefont {U.-J.}\ \bibnamefont
  {Wiese}},\ }\href@noop {} {\bibfield  {journal} {\bibinfo  {journal}
  {Physical review letters}\ }\textbf {\bibinfo {volume} {94}},\ \bibinfo
  {pages} {170201.1} (\bibinfo {year} {2005})}\BibitemShut {NoStop}%
\bibitem [{\citenamefont {Wu}\ and\ \citenamefont
  {Hsieh}(2019)}]{tfd_prl_2019}%
  \BibitemOpen
  \bibfield  {author} {\bibinfo {author} {\bibfnamefont {J.}~\bibnamefont
  {Wu}}\ and\ \bibinfo {author} {\bibfnamefont {T.~H.}\ \bibnamefont {Hsieh}},\
  }\href {https://doi.org/10.1103/PhysRevLett.123.220502} {\bibfield  {journal}
  {\bibinfo  {journal} {Phys. Rev. Lett.}\ }\textbf {\bibinfo {volume} {123}},\
  \bibinfo {pages} {220502} (\bibinfo {year} {2019})}\BibitemShut {NoStop}%
\bibitem [{\citenamefont {Zhu}\ \emph {et~al.}(2020)\citenamefont {Zhu},
  \citenamefont {Chang}, \citenamefont {Hsieh}, \citenamefont {Brown},
  \citenamefont {Linke},\ and\ \citenamefont {Monroe}}]{tfd_exp_pnas_2020}%
  \BibitemOpen
  \bibfield  {author} {\bibinfo {author} {\bibfnamefont {D.}~\bibnamefont
  {Zhu}}, \bibinfo {author} {\bibfnamefont {C.~S.}\ \bibnamefont {Chang}},
  \bibinfo {author} {\bibfnamefont {M.}~\bibnamefont {Hsieh}}, \bibinfo
  {author} {\bibfnamefont {K.~R.}\ \bibnamefont {Brown}}, \bibinfo {author}
  {\bibfnamefont {N.~M.}\ \bibnamefont {Linke}},\ and\ \bibinfo {author}
  {\bibfnamefont {C.}~\bibnamefont {Monroe}},\ }\href
  {https://doi.org/10.1073/pnas.2006337117} {\bibfield  {journal} {\bibinfo
  {journal} {Proc. Natl. Acad. Sci. USA}\ }\textbf {\bibinfo {volume} {117}},\
  \bibinfo {pages} {25402} (\bibinfo {year} {2020})}\BibitemShut {NoStop}%
\bibitem [{\citenamefont {Sagastizabal}\ \emph {et~al.}(2021)\citenamefont
  {Sagastizabal}, \citenamefont {Bonet-Monroig}, \citenamefont {Ding},
  \citenamefont {Egger}, \citenamefont {Barkoutsos}, \citenamefont
  {Ollitrault}, \citenamefont {Kandala}, \citenamefont {Temme}, \citenamefont
  {Gambetta},\ and\ \citenamefont {Tavernelli}}]{tfd_npjq_2021}%
  \BibitemOpen
  \bibfield  {author} {\bibinfo {author} {\bibfnamefont {R.}~\bibnamefont
  {Sagastizabal}}, \bibinfo {author} {\bibfnamefont {X.}~\bibnamefont
  {Bonet-Monroig}}, \bibinfo {author} {\bibfnamefont {Y.}~\bibnamefont {Ding}},
  \bibinfo {author} {\bibfnamefont {D.~J.}\ \bibnamefont {Egger}}, \bibinfo
  {author} {\bibfnamefont {P.}~\bibnamefont {Barkoutsos}}, \bibinfo {author}
  {\bibfnamefont {P.}~\bibnamefont {Ollitrault}}, \bibinfo {author}
  {\bibfnamefont {A.}~\bibnamefont {Kandala}}, \bibinfo {author} {\bibfnamefont
  {K.}~\bibnamefont {Temme}}, \bibinfo {author} {\bibfnamefont {J.~M.}\
  \bibnamefont {Gambetta}},\ and\ \bibinfo {author} {\bibfnamefont
  {I.}~\bibnamefont {Tavernelli}},\ }\href
  {https://doi.org/10.1038/s41534-021-00461-0} {\bibfield  {journal} {\bibinfo
  {journal} {npj Quantum Inf.}\ }\textbf {\bibinfo {volume} {7}},\ \bibinfo
  {pages} {130} (\bibinfo {year} {2021})}\BibitemShut {NoStop}%
\bibitem [{\citenamefont {Motta}\ \emph {et~al.}(2020)\citenamefont {Motta},
  \citenamefont {Sun}, \citenamefont {Tan}, \citenamefont {O'Rourke},
  \citenamefont {Ye}, \citenamefont {Minnich}, \citenamefont {Brand{\~a}o},\
  and\ \citenamefont {Chan}}]{qite_natphys_2020}%
  \BibitemOpen
  \bibfield  {author} {\bibinfo {author} {\bibfnamefont {M.}~\bibnamefont
  {Motta}}, \bibinfo {author} {\bibfnamefont {C.}~\bibnamefont {Sun}}, \bibinfo
  {author} {\bibfnamefont {A.~T.~K.}\ \bibnamefont {Tan}}, \bibinfo {author}
  {\bibfnamefont {M.~J.}\ \bibnamefont {O'Rourke}}, \bibinfo {author}
  {\bibfnamefont {E.}~\bibnamefont {Ye}}, \bibinfo {author} {\bibfnamefont
  {A.~J.}\ \bibnamefont {Minnich}}, \bibinfo {author} {\bibfnamefont {F.~G.
  S.~L.}\ \bibnamefont {Brand{\~a}o}},\ and\ \bibinfo {author} {\bibfnamefont
  {G.~K.-L.}\ \bibnamefont {Chan}},\ }\href
  {https://doi.org/10.1038/s41567-019-0704-4} {\bibfield  {journal} {\bibinfo
  {journal} {Nat. Phys.}\ }\textbf {\bibinfo {volume} {16}},\ \bibinfo {pages}
  {205} (\bibinfo {year} {2020})}\BibitemShut {NoStop}%
\bibitem [{\citenamefont {Sun}\ \emph {et~al.}(2021)\citenamefont {Sun},
  \citenamefont {Motta}, \citenamefont {Tazhigulov}, \citenamefont {Tan},
  \citenamefont {Chan},\ and\ \citenamefont {Minnich}}]{qite_prxq_2021}%
  \BibitemOpen
  \bibfield  {author} {\bibinfo {author} {\bibfnamefont {S.-N.}\ \bibnamefont
  {Sun}}, \bibinfo {author} {\bibfnamefont {M.}~\bibnamefont {Motta}}, \bibinfo
  {author} {\bibfnamefont {R.~N.}\ \bibnamefont {Tazhigulov}}, \bibinfo
  {author} {\bibfnamefont {A.~T.~K.}\ \bibnamefont {Tan}}, \bibinfo {author}
  {\bibfnamefont {G.~K.-L.}\ \bibnamefont {Chan}},\ and\ \bibinfo {author}
  {\bibfnamefont {A.~J.}\ \bibnamefont {Minnich}},\ }\href
  {https://doi.org/10.1103/PRXQuantum.2.010317} {\bibfield  {journal} {\bibinfo
   {journal} {PRX Quantum}\ }\textbf {\bibinfo {volume} {2}},\ \bibinfo {pages}
  {010317} (\bibinfo {year} {2021})}\BibitemShut {NoStop}%
\bibitem [{\citenamefont {Kamakari}\ \emph {et~al.}(2022)\citenamefont
  {Kamakari}, \citenamefont {Sun}, \citenamefont {Motta}, \citenamefont
  {Minnich},\ and\ \citenamefont {Chan}}]{qite_open_prxq_2022}%
  \BibitemOpen
  \bibfield  {author} {\bibinfo {author} {\bibfnamefont {H.}~\bibnamefont
  {Kamakari}}, \bibinfo {author} {\bibfnamefont {S.-N.}\ \bibnamefont {Sun}},
  \bibinfo {author} {\bibfnamefont {M.}~\bibnamefont {Motta}}, \bibinfo
  {author} {\bibfnamefont {A.~J.}\ \bibnamefont {Minnich}},\ and\ \bibinfo
  {author} {\bibfnamefont {G.~K.-L.}\ \bibnamefont {Chan}},\ }\href
  {https://doi.org/10.1103/PRXQuantum.3.010320} {\bibfield  {journal} {\bibinfo
   {journal} {PRX Quantum}\ }\textbf {\bibinfo {volume} {3}},\ \bibinfo {pages}
  {010320} (\bibinfo {year} {2022})}\BibitemShut {NoStop}%
\bibitem [{\citenamefont {Zhang}\ \emph {et~al.}(2021)\citenamefont {Zhang},
  \citenamefont {Zhang}, \citenamefont {Xue}, \citenamefont {Zhu},\ and\
  \citenamefont {Wang}}]{zhang2021continuous}%
  \BibitemOpen
  \bibfield  {author} {\bibinfo {author} {\bibfnamefont {D.-B.}\ \bibnamefont
  {Zhang}}, \bibinfo {author} {\bibfnamefont {G.-Q.}\ \bibnamefont {Zhang}},
  \bibinfo {author} {\bibfnamefont {Z.-Y.}\ \bibnamefont {Xue}}, \bibinfo
  {author} {\bibfnamefont {S.-L.}\ \bibnamefont {Zhu}},\ and\ \bibinfo {author}
  {\bibfnamefont {Z.}~\bibnamefont {Wang}},\ }\href@noop {} {\bibfield
  {journal} {\bibinfo  {journal} {Physical review letters}\ }\textbf {\bibinfo
  {volume} {127}},\ \bibinfo {pages} {020502} (\bibinfo {year}
  {2021})}\BibitemShut {NoStop}%
\bibitem [{\citenamefont {Martinez}\ \emph {et~al.}(2016)\citenamefont
  {Martinez}, \citenamefont {Muschik}, \citenamefont {Schindler}, \citenamefont
  {Nigg}, \citenamefont {Erhard}, \citenamefont {Heyl}, \citenamefont {Hauke},
  \citenamefont {Dalmonte}, \citenamefont {Monz}, \citenamefont {Zoller},\ and\
  \citenamefont {Blatt}}]{martinez_nature_2016}%
  \BibitemOpen
  \bibfield  {author} {\bibinfo {author} {\bibfnamefont {E.~A.}\ \bibnamefont
  {Martinez}}, \bibinfo {author} {\bibfnamefont {C.~A.}\ \bibnamefont
  {Muschik}}, \bibinfo {author} {\bibfnamefont {P.}~\bibnamefont {Schindler}},
  \bibinfo {author} {\bibfnamefont {D.}~\bibnamefont {Nigg}}, \bibinfo {author}
  {\bibfnamefont {A.}~\bibnamefont {Erhard}}, \bibinfo {author} {\bibfnamefont
  {M.}~\bibnamefont {Heyl}}, \bibinfo {author} {\bibfnamefont {P.}~\bibnamefont
  {Hauke}}, \bibinfo {author} {\bibfnamefont {M.}~\bibnamefont {Dalmonte}},
  \bibinfo {author} {\bibfnamefont {T.}~\bibnamefont {Monz}}, \bibinfo {author}
  {\bibfnamefont {P.}~\bibnamefont {Zoller}},\ and\ \bibinfo {author}
  {\bibfnamefont {R.}~\bibnamefont {Blatt}},\ }\href
  {https://doi.org/10.1038/nature18318} {\bibfield  {journal} {\bibinfo
  {journal} {Nature}\ }\textbf {\bibinfo {volume} {534}},\ \bibinfo {pages}
  {516} (\bibinfo {year} {2016})}\BibitemShut {NoStop}%
\bibitem [{\citenamefont {Monroe}\ \emph {et~al.}(2021)\citenamefont {Monroe},
  \citenamefont {Campbell}, \citenamefont {Duan}, \citenamefont {Gong},
  \citenamefont {Gorshkov}, \citenamefont {Hess}, \citenamefont {Islam},
  \citenamefont {Kim}, \citenamefont {Linke}, \citenamefont {Pagano},
  \citenamefont {Richerme}, \citenamefont {Senko},\ and\ \citenamefont
  {Yao}}]{monroe_rmp_2021}%
  \BibitemOpen
  \bibfield  {author} {\bibinfo {author} {\bibfnamefont {C.}~\bibnamefont
  {Monroe}}, \bibinfo {author} {\bibfnamefont {W.~C.}\ \bibnamefont
  {Campbell}}, \bibinfo {author} {\bibfnamefont {L.-M.}\ \bibnamefont {Duan}},
  \bibinfo {author} {\bibfnamefont {Z.-X.}\ \bibnamefont {Gong}}, \bibinfo
  {author} {\bibfnamefont {A.~V.}\ \bibnamefont {Gorshkov}}, \bibinfo {author}
  {\bibfnamefont {P.~W.}\ \bibnamefont {Hess}}, \bibinfo {author}
  {\bibfnamefont {R.}~\bibnamefont {Islam}}, \bibinfo {author} {\bibfnamefont
  {K.}~\bibnamefont {Kim}}, \bibinfo {author} {\bibfnamefont {N.~M.}\
  \bibnamefont {Linke}}, \bibinfo {author} {\bibfnamefont {G.}~\bibnamefont
  {Pagano}}, \bibinfo {author} {\bibfnamefont {P.}~\bibnamefont {Richerme}},
  \bibinfo {author} {\bibfnamefont {C.}~\bibnamefont {Senko}},\ and\ \bibinfo
  {author} {\bibfnamefont {N.~Y.}\ \bibnamefont {Yao}},\ }\href
  {https://doi.org/10.1103/RevModPhys.93.025001} {\bibfield  {journal}
  {\bibinfo  {journal} {Rev. Mod. Phys.}\ }\textbf {\bibinfo {volume} {93}},\
  \bibinfo {pages} {025001} (\bibinfo {year} {2021})}\BibitemShut {NoStop}%
\bibitem [{\citenamefont {Farrell}\ \emph {et~al.}(2024)\citenamefont
  {Farrell}, \citenamefont {Illa}, \citenamefont {Savage}, \citenamefont
  {Bedaque} \emph {et~al.}}]{prxq_2024_schwinger100}%
  \BibitemOpen
  \bibfield  {author} {\bibinfo {author} {\bibfnamefont {R.}~\bibnamefont
  {Farrell}}, \bibinfo {author} {\bibfnamefont {{\'A}.~M.}\ \bibnamefont
  {Illa}}, \bibinfo {author} {\bibfnamefont {M.~J.}\ \bibnamefont {Savage}},
  \bibinfo {author} {\bibfnamefont {P.}~\bibnamefont {Bedaque}}, \emph
  {et~al.},\ }\href {https://doi.org/10.1103/PRXQuantum.5.020315} {\bibfield
  {journal} {\bibinfo  {journal} {PRX Quantum}\ }\textbf {\bibinfo {volume}
  {5}},\ \bibinfo {pages} {020315} (\bibinfo {year} {2024})}\BibitemShut
  {NoStop}%
\bibitem [{\citenamefont {Zohar}\ \emph {et~al.}(2016)\citenamefont {Zohar},
  \citenamefont {Cirac},\ and\ \citenamefont {Reznik}}]{zohar_rpp_2016}%
  \BibitemOpen
  \bibfield  {author} {\bibinfo {author} {\bibfnamefont {E.}~\bibnamefont
  {Zohar}}, \bibinfo {author} {\bibfnamefont {J.~I.}\ \bibnamefont {Cirac}},\
  and\ \bibinfo {author} {\bibfnamefont {B.}~\bibnamefont {Reznik}},\ }\href
  {https://doi.org/10.1088/0034-4885/79/1/014401} {\bibfield  {journal}
  {\bibinfo  {journal} {Rep. Prog. Phys.}\ }\textbf {\bibinfo {volume} {79}},\
  \bibinfo {pages} {014401} (\bibinfo {year} {2016})}\BibitemShut {NoStop}%
\bibitem [{\citenamefont {Gross}\ and\ \citenamefont
  {Bloch}(2017)}]{gross_science_2017}%
  \BibitemOpen
  \bibfield  {author} {\bibinfo {author} {\bibfnamefont {C.}~\bibnamefont
  {Gross}}\ and\ \bibinfo {author} {\bibfnamefont {I.}~\bibnamefont {Bloch}},\
  }\href {https://doi.org/10.1126/science.aal3837} {\bibfield  {journal}
  {\bibinfo  {journal} {Science}\ }\textbf {\bibinfo {volume} {357}},\ \bibinfo
  {pages} {995} (\bibinfo {year} {2017})}\BibitemShut {NoStop}%
\bibitem [{\citenamefont {Huang}\ \emph {et~al.}(2020)\citenamefont {Huang},
  \citenamefont {Kueng},\ and\ \citenamefont {Preskill}}]{classical_shadow}%
  \BibitemOpen
  \bibfield  {author} {\bibinfo {author} {\bibfnamefont {H.-Y.}\ \bibnamefont
  {Huang}}, \bibinfo {author} {\bibfnamefont {R.}~\bibnamefont {Kueng}},\ and\
  \bibinfo {author} {\bibfnamefont {J.}~\bibnamefont {Preskill}},\ }\href
  {https://doi.org/10.1038/s41567-020-0932-7} {\bibfield  {journal} {\bibinfo
  {journal} {Nat. Phys.}\ }\textbf {\bibinfo {volume} {16}},\ \bibinfo {pages}
  {1050} (\bibinfo {year} {2020})}\BibitemShut {NoStop}%
\end{thebibliography}%
\end{document}